\documentclass[10pt]{article}
\usepackage{extsizes,graphicx,amssymb,amsmath,amsfonts,mathtext,color,wrapfig,ytableau}

\def\be{\begin{eqnarray}}
\def\ee{\end{eqnarray}}

\hoffset= - 0.5in         

\begin{document}

\thispagestyle{empty}

\baselineskip14pt

\ytableausetup{boxsize=5pt}

\hfill ITEP/TH-29/13

\bigskip

\bigskip

\centerline{\Large{Colored knot amplitudes and Hall-Littlewood polynomials
}}

\vspace{3ex}

\centerline{\large{\emph{Sh.Shakirov
\footnote{Department of Mathematics and BCTP, UC Berkeley, USA; shakirov@math.berkeley.edu \\ ITEP, Moscow, Russia; shakirov@itep.ru }}}}

\vspace{3ex}



\centerline{ABSTRACT}

\bigskip

{\footnotesize
The amplitudes of refined Chern-Simons (CS) theory, colored by antisymmetric (or symmetric) representations, conjecturally generate the $\Lambda^r$- (or $S^r$-) colored triply graded homology of $(n,m)$ torus knots. This paper is devoted to the generalization of Rosso-Jones formula to refined amplitudes, that involves non-trivial $\Gamma$-factors -- expansion coefficients in the Macdonald basis. We derive from refined CS theory a linear recursion w.r.t. transformations $(n,m) \mapsto (n, n+m)$ and $(n,m) \mapsto (m,-n)$ that fully determines these factors. Applying this recursion to $(n,nk+1)$ torus knots colored by antisymmetric representations $[1^r]$ we prove that their amplitudes are rectangular $[n^r]$ Hall-Littlewood polynomials under $k$ units of framing (a.k.a. the Bergeron-Garsia $\nabla$) operator. For symmetric representations $[r]$, we find the dual -- $q$-Whittaker -- polynomials. These results confirm and give a colored extension of the observation of arXiv:1201.3339 that triply graded homology of many torus knots has a strikingly simple description in terms of Hall-Littlewood polynomials.
}

\section{Introduction}

$SU(N)$, level $k$ CS theory \cite{ChernSimons, WittenJones, WittenLectures} associates to a knot $K$ and a $SU(N)$ representation of highest weight $R = [R_1 \geq R_2 \geq \ldots \geq R_{N-1} ]$ a number $Z_R(K)$, called $R$-colored knot amplitude. These amplitudes have attracted attention in knot theory ever since the work of Witten \cite{WittenJones}, who showed that for $R = \square$ and $N = 2$ the knot amplitude equals the famous knot invariant -- the Jones polynomial $J(\textbf{q})$ \cite{Jones} with parameter $\textbf{q} = \sqrt{q}$ where
\begin{align}
q = \exp\left( \frac{2 \pi i}{k + N} \right)
\end{align}
Furthermore, for $R = \square$ and general $N$, the knot amplitude reproduces more general knot invariants -- the HOMFLY polynomials $H(\textbf{q}, \textbf{a})$ \cite{Homfly} with
$\textbf{q} = \sqrt{q}$, $\textbf{a} = \textbf{q}^N$. Most generally, for general $R$ and general $N$, the knot amplitude gives $R$-colored HOMFLY polynomials. Thus CS theory provides a unifying framework for many different polynomial knot invariants, allowing to study them as a single entity. Moreover, it reveals links to other subjects, such as conformal field theory \cite{WittenJones, LecturesRCFT, LecturesCFT, Verlinde}, matrix models \cite{AKMV, Marino} and integrability \cite{RT, LecturesQG}.

Colored HOMFLY polynomials remained to be the most general polynomial knot invariants for some time -- until the work of Khovanov \cite{Khovanov1, Khovanov2, KhovanovRozansky}, who discovered that all the above classical polynomial knot invariants can be realized as Euler characteristics of certain complexes of vector spaces associated to knots. The homologies of these complexes, now known as \emph{knot homologies}, give rise to yet more general polynomial knot invariants. Passing from the HOMFLY polynomial to the Poincare polynomial of corresponding homology theory (the weighted sum of Betti numbers with weight $\textbf{t}$) one obtains a new polynomial knot invariant $P(\textbf{q},\textbf{t},\textbf{a})$ that gives back the Euler characteristic -- i.e. the HOMFLY polynomial -- at $\textbf{t} = -1$. Following the work \cite{Superpolynomial}, $P(\textbf{q},\textbf{t},\textbf{a})$ is now called the \emph{superpolynomial}.

\pagebreak

It is a natural question whether it is possible to construct a \emph{refined} version of CS theory, that is, a deformation of ordinary CS theory that contains more information about the knot homology spaces -- ideally, that has superpolynomials as its knot amplitudes. To construct such a deformation, one first needs to choose some definition of the ordinary knot amplitudes. A deformation with required properties was found in \cite{AS} for a particular definition, which is applicable only to torus knots $K = K_{n,m}$ with winding numbers $n,m$ along the two cycles of the torus:
\begin{align}
Z_R(K_{n,m}) = \big< \varnothing \big| S \ {\cal W}_{R,n,m} \big| \varnothing \big>
\label{CS-Torus}
\end{align}
Here $S$ and ${\cal W}_{R,n,m}$ are certain operators, acting on the finite-dimensional Hilbert space ${\cal H}_{N,k}$ of symmetric polynomials in $N$ variables of degree no greater than $k$ in every variable. The natural basis in this space is the basis $\big|s_R\big>$ of Schur polynomials, with the lowest degree state $\big|s_\varnothing\big> \equiv \big|\varnothing\big> \equiv 1$ being the constant polynomial. The most important role is played by operator $S$, which -- together with another operator $T$ -- provides representation of the generators $S = \left(\begin{smallmatrix} 0& 1\\ -1& 0 \end{smallmatrix} \right)$ and $T = \left(\begin{smallmatrix} 1& 0\\ 1& 1 \end{smallmatrix} \right)$ of $SL(2,{\mathbb Z})$ (the mapping class group of the torus) that satisfy $S^4 = 1$, $(ST)^3 = S^2$. The second ingredient, the operator ${\cal W}_{R,n,m}$, is called knot operator, because it encodes the dependence on the knot and the coloring representation $R$. Known explicit formulas for operators $S,T$ and ${\cal W}_{R,n,m}$ make it possible to compute any desired knot amplitude $Z_R(K_{n,m})$ as a function of $q$.

In \cite{AS} it was argued that to obtain a refined version of CS theory one should take a look at the Macdonald $q,t$-deformation of (\ref{CS-Torus}). The proposal of \cite{AS} was to leave the Hilbert space ${\cal H}_{N,k}$ intact, but replace the basis of Schur polynomials $\big| s_R \big>$ by Macdonald polynomials $\big| M_R \big>$ with parameters
\begin{align}
q = \exp\left( \frac{2 \pi i}{k + \beta N} \right), \
t = \exp\left( \frac{2 \pi i \beta}{k + \beta N} \right)
\label{rootsofunity}
\end{align}
for some nonzero $\beta \in {\mathbb C}^*$, and -- most importantly -- replace the usual $S$ and $T$ operators by their $\beta$-deformations, that satisfy $S^4 = 1$, $(ST)^3 = S^2$ identically for arbitrary value of $\beta$. Such operators have been known for a while in the framework of modular tensor categories \cite{MTC}; in \cite{AS} they have been independently derived from string theory. Again, known explicit formulas (Defs. 2.3 and 2.5 below) for the $\beta$-deformed operators $S,T$ and the knot operators ${\cal W}_{R,n,m}$ make it possible to compute any desired knot amplitude $Z_R(K_{n,m})$ as a function of $q$ and $t$.

It has been shown in \cite{AS}, later in \cite{AS2,MoscowHallLittlewood} and \cite{Cherednik} \footnote{in the independent framework of doubly affine Hecke algebras (DAHA), that has been later proved \cite{GammaNegut} to be equivalent to refined Chern-Simons theory.} that in many particular examples with $R$ being fundamental $\square = [1]$ or, more generally, symmetric $[r]$ or antisymmetric $[1^r]$ representation, such refined CS amplitudes coincide \footnote{ Note that, for other representations $R$ and/or torus links (with non-coprime $n,m$) refined CS amplitudes may still differ from superpolynomials. For more details about this important difference see \cite{Quadr, Zexpand}. } up to overall normalization with superpolynomials $P(\textbf{q},\textbf{t},\textbf{a})$, given that $\textbf{q} = \sqrt{t}, \textbf{t} = -\sqrt{q/t}, \textbf{a} = - \textbf{q}^N \textbf{t}^{-1}$. We will call this statement the main conjecture of refined CS theory. Among the other, it actually justifies the name "refined CS theory" and provides an efficient way to compute the superpolynomial in the above listed cases -- much simpler than to compute them by definition, via knot homology.

Refined CS theory described above is only applicable for torus knots. It would be interesting to find out, are there any generalizations of refined CS theory to non-torus knots. Developing such generalizations, if possible at all, may require better understanding of the structures that are already seen at the torus level. A particularly interesting structure that has been recently studied in \cite{GammaSmirnov, GammaOblomkov, GammaNegut, GammaGukov, MoscowHallLittlewood} is the refined analogue of the well-known Rosso-Jones formula \cite{RossoJones}:
\begin{align}
Z_R(K_{n,m}) = \sum\limits_{Y} M_Y(p^\star) \Gamma^{(n,m)}_Y
\label{GammaFactors}
\end{align}
where $M_Y(p^\star)$ are the Macdonald polynomials at the special point (the "topological locus")
\begin{align}
x^{\star}_i = t^{-i}, \ \ \ p^\star_k = \sum\limits_{i=1}^{N} \big(x^{\star}_i\big)^k = \dfrac{1 - t^{-Nk}}{1 - t^{-k}}
\end{align}
and $\Gamma^{(n,m)}_Y$ are certain coefficients. Just as in the ordinary Rosso-Jones formula, in this decomposition the basis polynomials $M_Y(p^\star)$ are the only quantities that depend on $N$, the coefficients $\Gamma^{(n,m)}_Y$ being $N$-independent. At $t = q$ these coefficients are equal to the Adams tensor product coefficients, while for $t \neq q$ such a simple interpretation is lost, as observed in \cite{GammaSmirnov}. Thus a natural question is to find a nice general expression for these $\Gamma$-factors.

Note that, from the viewpoint of refined CS theory, the Rosso-Jones decomposition (\ref{GammaFactors}) is nothing but decomposition of the matrix element (\ref{CS-Torus}) into intermediate states:
\begin{align}
Z_R(K_{n,m}) = \big< \varnothing \big| S \ {\cal W}_{R,n,m} \big| \varnothing \big> = \sum\limits_{Y} \underbrace{\big< \varnothing \big| S \big| M_Y \big>}_{M_Y(p^\star)} \underbrace{\big< M_Y \big| {\cal W}_{R,n,m} \big| \varnothing \big>}_{\Gamma^{(n,m)}_Y}
\end{align}
This identifies the $\Gamma$-factors with the matrix elements of knot operators of refined CS theory, and allows for their direct computation. In \cite{GammaNegut} a method was suggested that allows to compute these matrix elements, and an explicit expression was given for the case $R = \square$. Here we propose a method different from the one used in \cite{GammaNegut}, that allows to compute all the $\Gamma$-factors
\begin{align}
\Gamma^{(n,m)}_{R|A,B} = \big< M_A \big| {\cal W}_{R,n,m} \big| M_B \big>
\end{align}
with the help of linear relations \footnote{Note, that first of these relations was stated and actively used in \cite{GammaSmirnov} -- it is essentially the definition of \emph{evolution} of superpolynomials. The second relation can be thought to complement the evolution picture of \cite{GammaSmirnov} by describing the change under transformation $(n,m) \mapsto (m,-n)$. }
\begin{align}
\boxed{
T_A \Gamma^{(n,m)}_{R|A,B} = \Gamma^{(n,n+m)}_{R|A,B} T_B, \ \ \
\sum\limits_{Y} S_{A,Y} \Gamma^{(n,m)}_{R|Y,B} = \sum\limits_{Y} \Gamma^{(m,-n)}_{R|A,Y} S_{Y,B} }
\label{MainRecursion}
\end{align}
The paper is organized as follows. In \textbf{section 2} we recall the definition of refined CS theory, following \cite{AS}. In \textbf{section 3} we introduce the $\Gamma$-factors, derive the main recursion eq. (\ref{MainRecursion}) and explain how to solve it in practice using a technical tool called stable limit. In \textbf{Section 4} we solve it for $(n,nk+1)$ torus knots, colored by any antisymmetric $[1^r]$ representation, obtaining
\begin{align}
\boxed{
\Gamma^{(n,nk+1)}_{[1^r]\big|Y, \varnothing} = T_Y^{k} \cdot \mbox{ the coefficient of } M_Y \mbox{ in the expansion of } \mbox{HL}_{[n^r]}
}
\end{align}
where $HL$ stands for the Hall-Littlewood polynomial (Macdonald polynomial at $q=0$). We also give a dual formula for the case of symmetric representations in terms of dual Hall-Littlewood polynomials (Macdonald polynomials at $t=0$). The proof of these two formulas reduces to a new identity for symmetric functions; this proof will be given in \textbf{Appendix A}, written in collaboration with A.Borodin, I.Corwin and V.Gorin. In \textbf{Appendix B} we present a examples of $\Gamma$-factors computed by our method.

\section{Refined CS theory}

Let us briefly remind the key definitions behind refined CS theory. For a broader review and explanation of the physical meaning of these ingredients, see \cite{AS,AS2}.

\paragraph{Definition 2.1.} Let $\Lambda_{N}$ be the algebra of class functions on $SU(N)$ -- symmetric polynomials in $N$ variables $x_1, \ldots, x_N$ with a relation $x_1 \ldots x_N = 1$ -- equipped with a scalar product $\big< \cdot \big| \cdot \big>$ defined on homogeneous elements by the following integral
\begin{align}
\big< f \big| g \big> = \dfrac{1}{N!} \ \oint\limits_{|z_1| = 1} \dfrac{dz_1}{z_1}  \ldots \oint\limits_{|z_N| = 1} \dfrac{dz_N}{z_N} \ \prod\limits_{i \neq j} \psi\left( \dfrac{z_i}{z_j} \right) \ {\check f}( z_1, \ldots, z_N ) \ {\check g}\big( {\overline z}_1, \ldots, {\overline z}_N \big)
\label{scalarp}
\end{align}
and extended to arbitrary elements by linearity. Here $\check f$ denotes a representative of minimal nonnegative degree in the equivalence class of $f$, and $\psi$ denotes a function
\begin{align}
\psi(x) = \prod\limits_{m = 0}^{\infty} \dfrac{1 - q^m x}{1 - q^m t x} = 1 + \dfrac{t-1}{1-q} x + \dfrac{(t-1)(t-q)}{(1 - q)(1 - q^2)} x^2 + \dfrac{(t-1)(t-q)(t-q^2)}{(1 - q)(1 - q^2)(1 - q^3)} x^3 + \ldots
\label{psi}
\end{align}
for a pair of nonzero complex numbers $q,t \in {\mathbb C}^*$. This is a deformation (refinement) of the natural scalar product of class functions, the undeformed case being $q = t$ and $\psi(x) = 1 - x$.

\paragraph{Definition 2.2.} The unique basis in $\Lambda_{N}$, orthogonal w.r.t. the scalar product above and upper-triangularly related to the monomial basis, is called the Macdonald basis. Its elements -- the Macdonald polynomials -- are denoted $ \big| M_R \big> \in \Lambda_{N}$, labeled by partitions (Young diagrams) \footnote{Throughout the paper we will denote Young diagrams by uppercase Roman letters, as it is often done in topological string literature, and not by lowercase Greek letters, as usually preferred in symmetric functions theory.} of length less than $N$, i.e. $R = [R_1 \geq R_2 \geq \ldots \geq R_{N-1}]$, of degree $|R| = \sum_i R_i$. Normalization of Macdonald polynomials can be chosen in different ways; we use the form
\begin{align}
\big< M_R \big| M_{R^{\prime}} \big> = \delta_{R, R^{\prime}} \ G_R, \ \ \ \ \ G_R = \prod\limits_{i < j}^{N} \dfrac{\psi\big(t^{i-j}q^{R_j-R_i}\big)}{\psi\big(t^{j-i}q^{R_i-R_j}\big)}
\label{Macdonald}
\end{align}

\paragraph{Definition 2.3.} Let $S,T: \Lambda_{N} \rightarrow \Lambda_{N}$ be linear operators with the following matrices:

\begin{align}
\left\{
\begin{array}{lll}
S_{A,B} = \dfrac{1}{G_A} \big< M_A \big| S \big| M_B \big> = S_{\varnothing,\varnothing} \ \dfrac{1}{G_A} \ M_A\big( t^{\rho_1}, \ldots, t^{\rho_N} \big) M_B\big( t^{\rho_1} q^{A_1}, \ldots, t^{\rho_N} q^{A_N} \big) \\
\\
T_{A,B} = \dfrac{1}{G_A} \big< M_A \big| T \big| M_B \big> = \delta_{A,B} \ \sqrt{t}^{||A^T|| - N |A|} \sqrt{q}^{-||A|| + |A|^2/N}
\end{array}
\right.
\label{STformulas}
\end{align}
\smallskip\\
Here $\rho$ is a vector with components $\rho_i = (N+1)/2 - i$, $||A|| = \sum_i A_i^2$, and $S_{\varnothing,\varnothing}$ is a  normalization that we won't need in this paper.
Operators $S,T$ are remarkable for the following reason:

\paragraph{Theorem 2.1. Kirillov \cite{MTC}} Given positive integer $k$, let ${\cal H}_{N,k} \subset \Lambda_N$ denote the finite-dimensional subspace spanned by $\big|M_Y\big>$ with $Y_1 \leq k$. If $q$ and $t$ are specialized to
\begin{align*}
q = \exp\left( \frac{2 \pi i}{k + \beta N} \right),
t = \exp\left( \frac{2 \pi i \beta}{k + \beta N} \right), \ \ \ \beta \in {\mathbb C}^*
\end{align*}
then matrix elements of $S,T$ vanish beyond ${\cal H}_{N,k}$ and satisfy $S^4 = {\rm id}$ and $(S T)^3 = S^2$ on ${\cal H}_{N,k}$. This implies that they define a linear representation $\rho: SL(2, {\mathbb Z}) \mapsto \mbox{GL}\big( {\cal H}_{N,k} \big)$, by sending the generators of the group to $\rho\left(\begin{smallmatrix} 0& 1\\ -1& 0 \end{smallmatrix} \right) = S $ and $ \rho\left(\begin{smallmatrix} 1& 0\\ 1& 1 \end{smallmatrix} \right) = T$. In particular, this implies that on ${\cal H}_{N,k}$ operators $S$ and $T$ are invertible with $S^{-1} = S^3$ and $T^{-1} = S^3 T S T S$.

\paragraph{Definition 2.4.} Let ${\cal O}_R: \Lambda_N \rightarrow \Lambda_N$ be the multiplication operator
\begin{align}
{\cal O}_R \big| M_B \big> = \big| M_R M_B \big> = \sum\limits_{A} N^{A}_{R,B} \big| M_A \big>
\end{align}
with matrix elements $N^{A}_{R,B}$ being the $q,t$-Littlewood-Richardson coefficients. This operator is closely related to $SL(2, {\mathbb Z})$ operators via the formula first discovered \cite{Verlinde} for $q=t$ by Verlinde:

\paragraph{Theorem 2.2. Verlinde.} $S {\cal O}_R S^{-1}$ is diagonal in the basis of Macdonald polynomials on ${\cal H}_{N,k}$.

\paragraph{Proof.} It is easy to see that
\begin{align}
\big< M_A \big| S {\cal O}_R \big| M_B \big> = \big< M_A \big| S \big| M_R M_B \big> =
\end{align}
\begin{align}
= S_{\varnothing,\varnothing} M_A\big( t^{\rho_1}, \ldots, t^{\rho_N} \big) \Big( M_R M_B \Big)\big( t^{\rho_1} q^{A_1}, \ldots, t^{\rho_N} q^{A_N} \big) = S_{A, B} \ \dfrac{ S_{R,A} }{ S_{R, \varnothing} }
\end{align}
hence on ${\cal H}_{N,k}$ one has
\begin{align}
\big< M_A \big| S {\cal O}_R S^{-1} \big| M_B \big> = \delta_{A, B} \ \dfrac{ S_{R,A} }{ S_{R, \varnothing} }
\end{align}
which is indeed diagonal.

\paragraph{Definition 2.5.} Let ${\cal W}_{R,n,m}: {\cal H}_{N,k} \rightarrow {\cal H}_{N,k}$ be a linear operator
\begin{align}
{\cal W}_{R,n,m} = {\cal U}_{n,m} \ {\cal O}_R \ {\cal U}_{n,m}^{-1}, \ \ \ {\cal U}_{n,m} = \rho\left( \begin{array}{cc} n & a \\ m & b \end{array} \right)
\end{align}
for some $a,b$ such that $n b - m a = 1$. Let us show that operator ${\cal W}_{R,n,m}$ is well-defined, i.e. does not depend on a particular choice of $a,b$. Indeed, if one picks a different such pair ${\widetilde a}, {\widetilde b}$
\begin{align}
{\widetilde {\cal W}}_{R,n,m} = {\widetilde {\cal U}}_{n,m} \ {\cal O}_R \ {\widetilde {\cal U}}_{n,m}^{-1}, \ \ \ {\widetilde {\cal U}}_{n,m} = \rho\left( \begin{array}{cc} n & {\widetilde a} \\ m & {\widetilde b} \end{array} \right)
\end{align}
then it follows that vector $\big( {\widetilde a} - a , {\widetilde b} - b \big) = k \big(n,m\big)$ and hence ${\widetilde {\cal U}}_{n,m} = {\cal U}_{n,m} (TST)^k$. However, such a change of ${\cal U}_{n,m}$ does not affect ${\cal W}_{R,n,m}$ for the reason that ${\cal O}_R$ commutes with $TST$. Indeed, since operators $S {\cal O}_R S^{-1}$ and $T$ are both diagonal, they commute with each other:
\begin{align*}
T S {\cal O}_R S^{-1} = S {\cal O}_R S^{-1} T
\end{align*}
This implies that ${\cal O}_R$ commutes with $S^{-1} T^{-1} S$, which is the same as $TST$ via the $SL(2, {\mathbb Z})$ relation $S^{-1} T^{-1} S = T S T$. The equality ${\widetilde {\cal W}}_{R,n,m} = {\cal W}_{R,n,m}$ then follows, since the extra $TST$ factors can be commuted through ${\cal O}_R$ and cancel each other. So knot operators ${\cal W}_{R,n,m}$ are well-defined.

\paragraph{Definition 2.6. Refined CS amplitude} of $K_{n,m}$ is the following matrix element:
\begin{align}
Z_R(K_{n,m}) = \big< \varnothing \big| S \ {\cal W}_{R,n,m} \big| \varnothing \big>
\end{align}
This concludes section 2. Amplitudes $Z_R(K_{n,m})$ are the main object of study in refined CS theory. We now concentrate on the knot operators ${\cal W}_{R,n,m}$ and their matrix elements, called $\Gamma$-factors.

\pagebreak

\section{Gamma-factors and main recursion}

\paragraph{Definition 3.1. $\Gamma$-factors} are the matrix elements of knot operators:
\begin{align}
\Gamma^{(n,m)}_{R|A,B} = \big< M_A \big| {\cal W}_{R,n,m} \big| M_B \big>
\label{GammaFactorsDef}
\end{align}

\paragraph{Definition 3.2.} Note, that $\Gamma$-factors with $B = \varnothing$ are expansion coefficients of the amplitudes:
\begin{align*}
Z_R(K_{n,m}) = \big< \varnothing \big| S \ {\cal W}_{R,n,m} \big| \varnothing \big> = \sum\limits_{Y \in \Lambda_{N,k}} \Gamma^{(n,m)}_{R|Y,\varnothing} S_{\varnothing, Y} = \sum\limits_{Y \in \Lambda_{N,k}} \Gamma^{(n,m)}_{R|Y,\varnothing} M_Y( p^{\star} )
\end{align*}
Motivated by this, let us call the following vector
\begin{align}
\big| P^{(n,m)}_R \big> = \sum\limits_{Y} \underline{\Gamma}^{(n,m)}_{R|Y,\varnothing} \big| M_Y \big>
\label{Extended}
\end{align}
the {\bf extended} refined CS amplitude of $K_{n,m}$, in the sence that $\big| P^{(n,m)}_R \big>$ is the amplitude $Z_R(K_{n,m})$ extended from the topological locus specialization $p = p^{\star}$ to arbitrary specialization.

\paragraph{Theorem 3.1. Recursion.} The matrix elements $\Gamma^{(n,m)}_{R|A,B}$ satisfy a pair of linear relations
\begin{align}
T_A \Gamma^{(n,m)}_{R|A,B} = \Gamma^{(n,n+m)}_{R|A,B} T_B
\label{Recursion1}
\end{align}
\begin{align}
\mathop{\sum\limits_{l(Y) < N}}_{h(Y) \leq k}
S_{A,Y} \Gamma^{(n,m)}_{R|Y,B} =
\mathop{\sum\limits_{l(Y) < N}}_{h(Y) \leq k} \Gamma^{(m,-n)}_{R|A,Y} S_{Y,B}
\label{Recursion2}
\end{align}
where the sum is over Young diagrams corresponding to basis elements of ${\cal H}_{N,k}$, that is, of length $l(Y) = Y_1^T < N$ and height $h(Y) = Y_1 \leq k$. This is a finite set, but its size grows with $N$ and $k$.

\paragraph{Proof.} Being the representations of $SL(2,{\mathbb Z})$ elements, the operators ${\cal U}_{n,m}$ satisfy
\begin{align}
{\cal U}_{n,n+m} = T {\cal U}_{n,m}, \ \ \ {\cal U}_{m,-n} = S {\cal U}_{n,m}
\end{align}
Therefore,
\begin{align}
{\cal W}_{R,n,n+m} T = T {\cal W}_{R,n,m}, \ \ \ {\cal W}_{R,m,-n} S = S {\cal W}_{R,n,m}
\end{align}
Written in terms of matrix elements, this is equivalent to the statement of the theorem.

\paragraph{} To solve the main recursion in practice, we will use the following two properties of $\Gamma$-factors. These properties are by now well known and follow from the fact that operators ${\cal W}$ form a specific algebra, namely the spherical DAHA a.k.a. the shuffle algebra, as explained in \cite{GammaNegut}.

\paragraph{Property I. Degree.} $\Gamma^{(n,m)}_{R|A,B}$ vanishes unless $|A|-|B|-n|R| = 0 \mod N$.

\paragraph{Property II. Stabilization.} The $\Gamma$-factors with empty last index are stable, i.e. for sufficiently large $N,k$ each of these $\Gamma$-factors becomes equal (up to a normalization factor) to its stable limit
\begin{align}
\Gamma^{(n,m)}_{R|A,\varnothing} = {\rm const} \ \underline{\Gamma}^{(n,m)}_{R|A,\varnothing}
\end{align}
where for any function $f(N,k,\beta)$ we define its stable limit $\underline{f}(q,t)$ to be
\begin{align}
\underline{f}\big(q,t\big)
:= \lim\limits_{k,N \rightarrow \infty} f\left( e^{\frac{2\pi i}{k + \beta N}} = \mbox{fixed} = q, \ e^{\frac{2\pi i \beta}{k + \beta N}} = \mbox{fixed} = t \right)
\end{align}
and ${\rm const}$ is an overall normalization factor that depends on $R,n,m$ but is the same for all $A$.

\pagebreak

It is easy to find that the Macdonald norm $G$ and operators $S,T$ have stable limits, given by
\begin{align}
\underline{G}_B = \prod\limits_{(i,j) \in B} \dfrac{ 1 - t^{Arm_B(i,j) + 1} q^{Leg_B(i,j)} }{ 1 - t^{Arm_B(i,j) } q^{Leg_B(i,j) + 1} }
\label{StableG}
\end{align}
\begin{align}
\underline{S}_{A,B} = \underline{G}_A^{-1} M_A\left( t^{\rho} \right) M_B\left( t^{\rho} q^{A} \right)
\label{StableS}
\end{align}
\begin{align}
\underline{T}_{A,B} = \delta_{A,B} \ t^{||A^T||/2} q^{-||A||/2}
\label{StableT}
\end{align}
where $M_B\left( t^{\rho} q^{A} \right)$ stands for the specialization of the Macdonald symmetric function $M_B$ at
\begin{align}
p_k\big(t^{\rho} q^{A}\big) = \sum\limits_{i = 1}^{l(A)} \big( t^{-i} q^{A_i} \big)^k - \dfrac{t^{-k l(A)}}{1-t^k}
\end{align}
Note that, for the sake of simplicity, we slightly abused the notation by omitting the constant $\underline{S}_{\varnothing, \varnothing}$ in the r.h.s. of eq.(\ref{StableS}); this means that $\underline{S}_{A,B}$ stands here for the stable limit of $S_{A,B}/S_{\varnothing, \varnothing}$.

\paragraph{Theorem 3.2. Stable recursion.} The stable limits  $\underline{\Gamma}^{(n,m)}_{R|A,B}$ satisfy a pair of linear relations
\begin{align}
\underline{T}_A \underline{\Gamma}^{(n,m)}_{R|A,B} = \underline{\Gamma}^{(n,n+m)}_{R|A,B} \underline{T}_B
\label{StableRecursion1}
\end{align}
\begin{align}
\sum\limits_{|Y|=|B|+n|R|} \underline{S}_{A,Y} \underline{\Gamma}^{(n,m)}_{R|Y,B} = \sum\limits_{|Y|=|A|-m|R|} \underline{\Gamma}^{(m,-n)}_{R|A,Y} \underline{S}_{Y,B}
\label{StableRecursion2}
\end{align}

\paragraph{Proof.} Taking the stable limit of both sides of eq.(\ref{Recursion1}), one obtains eq. (\ref{StableRecursion1}):
\begin{align*}
T_A \Gamma^{(n,m)}_{R|A,B} = \Gamma^{(n,n+m)}_{R|A,B} T_B
\end{align*}
\begin{align*}
\downarrow \mbox{stable limit}
\end{align*}
\begin{align*}
\underline{T}_A \underline{\Gamma}^{(n,m)}_{R|A,B} = \underline{\Gamma}^{(n,n+m)}_{R|A,B} \underline{T}_B
\end{align*}
Taking the stable limit of both sides of eq.(\ref{Recursion2}), one obtains
\begin{align}
\mathop{\sum\limits_{l(Y) < N}}_{h(Y) \leq k}
S_{A,Y} \Gamma^{(n,m)}_{R|Y,B} =
\mathop{\sum\limits_{l(Y) < N}}_{h(Y) \leq k} \Gamma^{(m,-n)}_{R|A,Y} S_{Y,B}
\end{align}
\begin{align*}
\downarrow \mbox{stable limit}
\end{align*}
\begin{align}
\sum\limits_{|Y|=0}^{\infty} \underline{S}_{A,Y} \underline{\Gamma}^{(n,m)}_{R|Y,B} = \sum\limits_{|Y|=0}^{\infty} \underline{\Gamma}^{(m,-n)}_{R|A,Y} \underline{S}_{Y,B}
\end{align}
\begin{align*}
\downarrow \mbox{Property I}
\end{align*}
\begin{align}
\sum\limits_{|Y|=|B|+n|R|} \underline{S}_{A,Y} \underline{\Gamma}^{(n,m)}_{R|Y,B} = \sum\limits_{|Y|=|A|-m|R|} \underline{\Gamma}^{(m,-n)}_{R|A,Y} \underline{S}_{Y,B}
\end{align}
After the first step, the sum in the both sides is over all Young diagrams. On the second step we use Property I, which implies that the stable limit  $\underline{\Gamma}^{(n,m)}_{R|A,B}$ vanishes unless $|A|-|B|-n|R| = 0$. In the end, the sums in the both sides are over diagrams of fixed finite size, as in eq. (\ref{StableRecursion2}).

\paragraph{} Eqs. (\ref{StableRecursion1}) and (\ref{StableRecursion2}) provide an overdetermined system of linear equations for the unknowns $\underline{\Gamma}^{(n,m)}_{R|A,B}$ that completely determine them all. Since, by Property II, $\Gamma$-factors $\Gamma^{(n,m)}_{R|A,\varnothing}$ are essentially the same as their stable limits $\underline{\Gamma}^{(n,m)}_{R|A,\varnothing}$, this completely determines the former and, hence, determines the refined CS amplitudes for all torus knots. This is best seen at example:

\paragraph{Example: the trefoil, i.e. the (2,3) knot.}Let us provide a simple example of how the stable recursion works by using it to determine the $\Gamma$-factors of the $(2,3)$ torus knot, known as the trefoil, in the fundamental representation. The refined CS amplitude of this knot has a form
\begin{align*}
Z_\square(K_{2,3}) = \big< \varnothing \big| S \ {\cal W}_{\square,2,3} \big| \varnothing \big> \ \sim \
M_2(p^\star) \ \underline{\Gamma}^{(2,3)}_{\square |[2], \varnothing} + M_{11}(p^\star) \ \underline{\Gamma}^{(2,3)}_{\square |[11], \varnothing}
\end{align*}
where we used Property I to fix the size of diagrams and Property II to write \footnote{Note, that overall normalization cannot be computed by this method unless anything is said about the constant of proportionality in Property II. For the purpose of applications to superpolynomials, however, this imposes no difficulty, since the overall normalization of a superpolynomial is rarely of much interest.} the stable $\Gamma$-factors in place of the matrix elements of the knot operator.

\begin{wrapfigure}{l}{0.25\textwidth}
\vspace{-2ex}
  \begin{center}
    \includegraphics[width=0.25\textwidth]{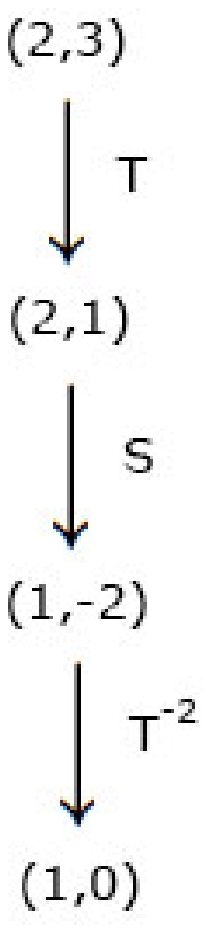}
  \end{center}
  \vspace{-2ex}
  \caption{The recursion path for the trefoil knot.}
\end{wrapfigure} The recursion follows the scheme depicted on Figure 1. Eq. (\ref{StableRecursion1}) can be used to transform the (2,3) knot into the (2,1) knot:
\begin{align*}
\underline{\Gamma}^{(2,3)}_{\square |[2], \varnothing} = \underline{T}_{2} \ \underline{\Gamma}^{(2,1)}_{\square |[2], \varnothing}, \ \ \ \underline{\Gamma}^{(2,3)}_{\square |[11], \varnothing} = \underline{T}_{11} \ \underline{\Gamma}^{(2,1)}_{\square |[11], \varnothing}
\end{align*}
Eq. (\ref{StableRecursion2}) can be used to transform the (2,1) knot into the (1,-2) knot:
\begin{align}
\left\{
\begin{array}{lll}
\underline{S}_{[2],[2]} \ \underline{\Gamma}^{(2,1)}_{\square |[2], \varnothing} + \underline{S}_{[2],[11]} \ \underline{\Gamma}^{(2,1)}_{\square |[11], \varnothing} = \underline{\Gamma}^{(1,-2)}_{\square |[2], [1]} \ \underline{S}_{[1], \varnothing} \\
\\
\underline{S}_{[11],[2]} \ \underline{\Gamma}^{(2,1)}_{\square |[2], \varnothing} + \underline{S}_{[11],[11]} \ \underline{\Gamma}^{(2,1)}_{\square |[11], \varnothing} = \underline{\Gamma}^{(1,-2)}_{\square |[11], [1]} \ \underline{S}_{[1], \varnothing}
\end{array}
\right.
\label{ExampleRecursion}
\end{align}
Eq. (\ref{StableRecursion1}) can be used to transform the (1,-2) knot into the (1,0) knot:
\begin{align*}
\underline{\Gamma}^{(1,-2)}_{\square |[2], [1]} = \left(\dfrac{\underline{T}_{2}}{\underline{T}_{1}}\right)^{-2} \ \underline{\Gamma}^{(1,0)}_{\square |[2], [1]}, \ \ \ \underline{\Gamma}^{(1,-2)}_{\square |[11], [1]} = \left(\dfrac{\underline{T}_{2}}{\underline{T}_{1}}\right)^{-2} \ \underline{\Gamma}^{(1,0)}_{\square |[11], [1]}
\end{align*}
This is the endpoint of the recursion: by definition, for the (1,0) knot the $\Gamma$-factors are nothing but the $q,t$-Littlewood-Richardson multiplication coefficients for the Macdonald polynomials: $\underline{\Gamma}^{(1,0)}_{R | A, B} = N_{R,B}^{A}$. Hence,
\begin{align*}
\underline{\Gamma}^{(1,0)}_{\square |[2], [1]} =
N_{[1], [1]}^{[2]} = 1, \ \ \ \underline{\Gamma}^{(1,0)}_{\square |[11], [1]} =
N_{[1], [1]}^{[11]} = \dfrac{(1-q)(1+t)}{1-qt}
\end{align*}
\begin{align*}
\underline{\Gamma}^{(1,-2)}_{\square |[2], [1]} = \dfrac{t}{q^3}, \ \ \ \underline{\Gamma}^{(1,-2)}_{\square |[11], [1]} = \dfrac{t^3}{q} \dfrac{(1-q)(1+t)}{1-qt}
\end{align*}
This, in turn, serves as a free term for the square linear system (\ref{ExampleRecursion}). Substituting all the required values of the ${\underline S}$-matrix and solving the square linear system, one finds for the (2,1) knot
\begin{align*}
\underline{\Gamma}^{(2,1)}_{\square |[2], \varnothing} = t, \ \ \ \underline{\Gamma}^{(2,1)}_{\square |[11], \varnothing} = tq \dfrac{t^2-1}{1-tq}
\end{align*}
and, finally, for the (2,3) knot
\begin{align*}
\underline{\Gamma}^{(2,3)}_{\square |[2], \varnothing} = \dfrac{t^2}{q^2}, \ \ \ \underline{\Gamma}^{(2,3)}_{\square |[11], \varnothing} = t^3 \dfrac{t^2-1}{(1-tq)}
\end{align*}
what is known to be a correct answer \cite{GammaSmirnov} up to an overall normalization.

\pagebreak

\section{$(n,nk+1)$ knots and Hall-Littlewood polynomials}

\begin{wrapfigure}{r}{0.3\textwidth}
\vspace{-8ex}
  \begin{center}
    \includegraphics[width=0.3\textwidth]{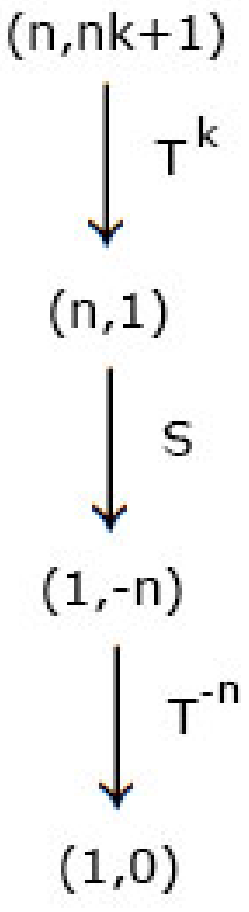}
  \end{center}
  \vspace{-2ex}
  \caption{The recursion path for the $(n,nk+1)$ knot.}
\end{wrapfigure}
Having demonstrated that the stable recursion indeed allows to compute the $\Gamma$-factors, let us now proceed to solve it in a more general case, the case of $(n,nk+1)$ knots. Note that, from the point of view of the recursion, there is not much difference between the case $(2,3)$ and $(n,nk+1)$. Indeed, as Figure 2 shows, the recursion schemes for these two knots are essentially identical and involve only one swap of the winding numbers, realized by the ${\underline S}$-matrix. The only difference is that inversion of the involved finite-dimensional block of the ${\underline S}$-matrix becomes significantly harder. Instead of simply inverting a huge matrix, it is desirable to find a conceptually smarter solution.

By Property I, the refined CS amplitude for the $(n,nk+1)$ knot is a sum over all Young diagrams of size $n |R|$:
\begin{align*}
Z_R(K_{n,nk+1}) \sim \
\sum\limits_{|Y|=n|R|} M_Y(p^\star) \ \underline{T}_{Y}^k \ \underline{\Gamma}^{(n,1)}_{R |Y, \varnothing}
\end{align*}
where we already used eq. (\ref{StableRecursion1}) to transform the $(n,nk+1)$ knot into the $(n,1)$ knot. The $\Gamma$-factors for the $(n,1)$ knot then have to be found, and the main equation for them is of course eq. (\ref{StableRecursion2}) that relates them to the $(1,-n)$ knot:
\begin{align*}
\sum\limits_{|Y|=n|R|} \underline{S}_{A,Y} \underline{\Gamma}^{(n,1)}_{R|Y,\varnothing} = \sum\limits_{|Y|=|A|-|R|} \underline{\Gamma}^{(1,-n)}_{R|A,Y} \underline{S}_{Y,\varnothing}
\end{align*}
Expressing the ${\underline S}$-matrices through Macdonald polynomials and the $\Gamma$-factors in the r.h.s. through the $q,t$-Littlewood-Richardson coefficients, this equation (system of equations) takes form
\begin{align}
\sum\limits_{|Y|=n|R|} M_Y(t^{\rho}q^{A}) \underline{\Gamma}^{(n,1)}_{R|Y,\varnothing} = \sum\limits_{|Y|=|A|-|R|} N_{R,Y}^{A} \ \left( \dfrac{\underline{T}_{A}}{\underline{T}_{Y}} \right)^{-n} \ \dfrac{M_Y(t^{\rho})}{M_A(t^{\rho})} \ \dfrac{{\underline G}_A}{{\underline G}_Y}
\label{nk+1}
\end{align}
In principle, one could proceed by inverting the finite square matrix in the l.h.s.; however, as we already emphasized, this is not the most elegant way out. It turns out that the nicest way out is to analyze the structure of the r.h.s., what makes solution of this system completely transparent, at least for completely symmetric or completely antisymmetric representations:

\paragraph{Theorem 4.1.} For $R = [1^r]$, the r.h.s. of eq. $(\ref{nk+1})$ is a Hall-Littlewood polynomial:
\begin{align}
\sum\limits_{|Y|=|A|-r} N_{[1^r],Y}^{A} \ \left( \dfrac{\underline{T}_{A}}{\underline{T}_{Y}} \right)^{-n} \ \dfrac{M_Y(t^{\rho})}{M_A(t^{\rho})} \ \dfrac{{\underline G}_A}{{\underline G}_Y} = {\rm const}_{n,r} \ \mbox{HL}_{[n^r]}(t^{\rho}q^{A})
\label{thm41}
\end{align}
where $\mbox{HL}$ means the Hall-Littlewood polynomial
\begin{align}
\mbox{HL}_{Y}\big\{p_k\big\} = M_{Y}\big\{p_k\big\} \Big|_{q=0}
\end{align}
and ${\rm const}_{n,r} = t^{r(r-2)/2} q^{-r(n-2)/2}$.

\paragraph{Theorem 4.2.} For $R = [r]$, the r.h.s. of eq. $(\ref{nk+1})$ is a dual Hall-Littlewood polynomial:
\begin{align}
\sum\limits_{|Y|=|A|-r} N_{[r],Y}^{A} \ \left( \dfrac{\underline{T}_{A}}{\underline{T}_{Y}} \right)^{-n} \ \dfrac{M_Y(t^{\rho})}{M_A(t^{\rho})} \ \dfrac{{\underline G}_A}{{\underline G}_Y} = {\rm const}^{\prime}_{n,r} \ \mbox{HL}^{\prime}_{[r^n]}(t^{\rho}q^{A})
\end{align}
where $\mbox{HL}^{\prime}$ means the (certain reparametrization of) the $q$-Whittaker polynomial
\begin{align}
\mbox{HL}^{\prime}_{Y}\big\{p_k\big\} = M_{Y}\big\{ (t^k - 1) p_k\big\} \Big|_{t = 0, q = q^{-1}}
\end{align}
and ${\rm const}_{n,r} = (-1)^{nr} t^{nr/2} q^{-r(n-2)/2} \prod\limits_{i=0}^{r-1} (1 - t q^i)^{-1} $.

\paragraph{} To the best of our knowledge, these are previously unknown identities in symmetric function theory. We devote a separate \textbf{Appendix A} to their proof, written in collaboration with A.Borodin, I.Corwin and V.Gorin. As a direct corollary of these identities, the $\Gamma$-factors for $(n,1)$ knots are nothing but Macdonald coefficients of polynomials $\mbox{HL}$ and $\mbox{HL}^{\prime}$:
\begin{align}
\underline{\Gamma}^{(n,1)}_{[1^r]|Y,\varnothing} = \dfrac{1}{||M_Y||^2} \Big< M_Y \Big| \mbox{HL}_{[n^r]} \Big>
\end{align}
\begin{align}
\underline{\Gamma}^{(n,1)}_{[r]|Y,\varnothing} = \dfrac{1}{||M_Y||^2} \Big< M_Y \Big| \mbox{HL}^{\prime}_{[r^n]} \Big>
\end{align}
where we omitted the overall constants of proportionality. These formulas generalize the explicit expressions of \cite{GammaGukov} from $n = 2$ to arbitrary $n$. For the extended amplitudes, one obtains
\begin{align}
\big| P^{(n,nk+1)}_{[1^r]} \big> = {\hat T}^k \ \Big| \mbox{HL}_{[n^r]} \Big>
\label{SuperA}
\end{align}
\begin{align}
\big| P^{(n,nk+1)}_{[r]} \big> = {\hat T}^k \ \Big| \mbox{HL}^{\prime}_{[r^n]} \Big>
\label{SuperS}
\end{align}
where, we remind, ${\hat T}$ is the framing operator that has eigenfunctions $\big|M_Y\big>$ with eigenvalues $T_Y$.

\section{Conclusion and discussion}

In this paper we considered a problem of calculation of the $\Gamma$-factors -- expansion coefficients of $R$-colored refined Chern-Simons amplitudes in Macdonald basis. This problem has recently attracted considerable attention, and in particular an explicit formula was presented in \cite{GammaNegut} for the case $R = \square$, the fundamental representation. We proposed a linear recursion in the winding numbers that computes the colored $\Gamma$-factors. For $(n,m) = (n,nk+1)$ and $R = [1^r]$ or $[r]$, we demonstrated that its solution is given by Hall-Littlewood or q-Whittaker polynomials, resp. This result is not isolated and is closely related to several research directions.

\paragraph{Hall-Littlewood polynomials and torus knots.} The fact that torus knot invariants often have elegant description in terms of Hall-Littlewood polynomials has been noticed in \cite{MoscowHallLittlewood}. This was pushed further in \cite{MoscowHallLittlewood2} for the unrefined ($t=q$, ordinary Chern-Simons theory) case, where it was conjectured that HOMFLY polynomial of arbitrary torus knot is precisely a Hall-Littlewood polynomial. This conjecture has been later proved in \cite{HallLittlewoodGorsky}. Eqs. (\ref{SuperA}),(\ref{SuperS}) provide a colored generalization of the original formulas of \cite{MoscowHallLittlewood} and at the same time a refined generalization of \cite{MoscowHallLittlewood2}.

\paragraph{Combinatorics of Macdonald polynomials.} It is important to note that the framing operator $T$ is the same as operator ${\hat \nabla}$, introduced by F.Bergeron and A.Garcia \cite{BergeronGarsia1, BergeronGarsia2}. They discovered that ${\hat \nabla}e_n$ is the Frobenius series for the ring of diagonal coinvariants, in particular, that in appropriate basis it has positive integer coefficients. The r.h.s. of eqs. (\ref{SuperA}),(\ref{SuperS}) provide a "colored" generalization of this identity, with ${\hat \nabla}e_n$ replaced by ${\hat \nabla}\mbox{HL}_{[n^r]}$, whose coefficients in appropriate basis have a meaning of Betty numbers for the triply graded knot homology, hence possess positivity and integrality.

\paragraph{Super-A-polynomials.} Though we made use of the recursion that acts on winding numbers, the results can be useful to study different other kinds of recursions, in particular, the super-A-polynomial recursion that was developed in \cite{GammaGukov} and acts on the spin (coloring representation $R$). Eqs. (\ref{SuperA}),(\ref{SuperS}) allow one to compute the super-A-polynomials for various $(n,nk+1)$ knots.

\section*{Acknowledgements}

We are grateful to E.Gorsky, S.Gukov, A.Mironov, A.Morozov for illuminating discussions and criticism. We are grateful to A.Borodin, I.Corwin and V.Gorin, in collaboration with whom a crucial part of this work, namely, the proof of the main identity for Hall-Littlewood polynomials, was written. This work was supported in part by grant RFBR 13-02-00525, grant for support of scientic schools NSH-3349.2012.2, and government contract 8606.

\pagebreak

\section*{Appendix A: Proofs}

To appear.

\section*{Appendix B: Extended amplitudes}

In this Appendix we present several explicit examples of $\Gamma$-factors (in the form of the extended amplitudes $\big| P^{(n,m)}_{R} \big>$ as defined in section 3) for various torus knots. Since uncolored $\Gamma$-factors are described in full in \cite{GammaNegut} with detailed examples given in the tables in \cite{GammaSmirnov}, we concentrate on going further into the colored realm. We present answers for the trefoil series $(2,2k+1)$ colored by arbitrary-shape representations $R$ with up to 5 boxes; $(3,3k+1)$ knots up to 4 and $(4,4k+1)$ knots up to 3 boxes. For each series, we give the extended amplitude of the simplest representative of the series, i.e. with winding numbers (2,1), (3,1) and (4,1), respectively, since the higher knots in the series can be easily obtained by acting $k$ times by the framing operator ${\hat T}$.

As noted in \cite{MoscowHallLittlewood}, it is often convenient to express the extended amplitudes in various other bases, different from the Macdonald basis. In this Appendix, we chose to present the extended amplitudes in the basis of modified Schur functions defined as
\begin{align}
{\widetilde S}_{Y}\big\{p_k\big\} = \dfrac{1}{1-t} S_{Y}\big\{ (1-t^k) p_k\big\}
\end{align}
The following table illustrates the various bases of symmetric functions used in present paper:
\begin{align*}
\begin{array}{c|c|c|c|c}
\mbox{Basis} & \mbox{Notation} & Y = [1] & Y = [11] & Y = [2] \\
\hline & & & & \\
\mbox{Macdonald} & M_Y & p_1 & \dfrac{1}{2} p_1^2 - \dfrac{1}{2} p_2 & \dfrac{(1+q)(1-t)}{2(1-qt)} p_1^2 + \dfrac{(1-q)(1+t)}{2(1-qt)} p_2 \\
& & & & \\
\mbox{Schur} & S_Y & p_1 & \dfrac{1}{2} p_1^2 - \dfrac{1}{2} p_2 & \dfrac{1}{2} p_1^2 + \dfrac{1}{2} p_2 \\
& & & & \\
\mbox{Hall-Littlewood} & {\rm HL}_Y & p_1 & \dfrac{1}{2} p_1^2 - \dfrac{1}{2} p_2 & \dfrac{1-t}{2} p_1^2 + \dfrac{1+t}{2} p_2 \\
& & & & \\
\mbox{Modified Schur} & {\widetilde S}_Y & p_1 & \dfrac{1-t}{2} p_1^2 - \dfrac{1+t}{2} p_2 & \dfrac{1-t}{2} p_1^2 + \dfrac{1+t}{2} p_2 \\
\end{array}
\end{align*}
As noted in \cite{MoscowHallLittlewood}, the benefit of using this basis is that all the coefficients of the extended amplitudes $\big| P^{(n,m)}_{R} \big>$ are polynomials in $q,t^{-1}$ (for simplicity in this section we denote $\tau = t^{-1}$), while in Macdonald basis they are rational functions. Moreover, in all the examples these coefficients are positive integer polynomials, suggesting an existence of some combinatorial interpretation.

\subsection*{Trefoil series (2,2k+1)}

\subsubsection*{Fundamental representation}

\begin{align*}
\big| P^{(2,1)}_{[1]} \big> = \big| {\widetilde S}_{2} \big>
\end{align*}

\subsubsection*{Representations of size $|R| = 2$}

\begin{align*}
\big| P^{(2,1)}_{[2]} \big> = \big| {\widetilde S}_{4} \big> + \tau \big| {\widetilde S}_{31} \big> + \tau^{2} \big| {\widetilde S}_{22} \big>
\end{align*}

\begin{align*}
\big| P^{(2,1)}_{[11]} \big> = \big| {\widetilde S}_{4} \big> + q \big| {\widetilde S}_{31} \big> + q^2 \big| {\widetilde S}_{22} \big>
\end{align*}

\subsubsection*{Representations of size $|R| = 3$}

\begin{align*}
\nonumber \big| P^{(2,1)}_{[111]} \big> =
\big| {\widetilde S}_{6} \big>
+ ( \tau^{2} + \tau ) \big| {\widetilde S}_{51} \big>
+ ( \tau^{2} + \tau^{3} + \tau^{4} ) \big| {\widetilde S}_{42} \big> \\
+ \tau^{3} \big| {\widetilde S}_{33} \big>
+ \tau^{3} \big| {\widetilde S}_{411} \big>
+ ( \tau^{4} + \tau^{5} ) \big| {\widetilde S}_{321} \big>
+ \tau^{6} \big| {\widetilde S}_{222} \big>
\end{align*}

\begin{align*}
\nonumber \big| P^{(2,1)}_{[21]} \big> =
\big| {\widetilde S}_{6} \big>
+ ( q + \tau ) \big| {\widetilde S}_{51} \big>
+ ( q^2 + q \tau + \tau^{2} ) \big| {\widetilde S}_{42} \big> \\
+ q \tau \big| {\widetilde S}_{33} \big>
+ q \tau \big| {\widetilde S}_{411} \big>
+ ( q \tau^{2} + q^2 \tau ) \big| {\widetilde S}_{321} \big>
+ q^2 \tau^{2} \big| {\widetilde S}_{222} \big>
\end{align*}

\begin{align*}
\nonumber \big| P^{(2,1)}_{[3]} \big> =
\big| {\widetilde S}_{6} \big>
+ ( q^{2} + q ) \big| {\widetilde S}_{51} \big>
+ ( q^{4} + q^{3} + q^{2} ) \big| {\widetilde S}_{42} \big> \\
+ q^{3} \big| {\widetilde S}_{33} \big>
+ q^{3} \big| {\widetilde S}_{411} \big>
+ ( q^{5} + q^{4} ) \big| {\widetilde S}_{321} \big>
+ q^{6} \big| {\widetilde S}_{222} \big>
\end{align*}

\subsubsection*{Representations of size $|R| = 4$}

\begin{center}
$\big| P^{(2,1)}_{[1111]} \big> =
 \big| {\widetilde S}_{8} \big>+(\tau+\tau^{2}+\tau^{3}) \big| {\widetilde S}_{7  1} \big>+(\tau^{2}+\tau^{3}+2 \tau^{4}+\tau^{5}+\tau^{6}) \big| {\widetilde S}_{6  2} \big>+(\tau^{3}+\tau^{4}+\tau^{5}) \big| {\widetilde S}_{6  1  1} \big> + (\tau^{3}+\tau^{4}+2 \tau^{5}+\tau^{6}+\tau^{7}) \big| {\widetilde S}_{5  3} \big>+(\tau^{4}+2 \tau^{5}+2 \tau^{6}+2 \tau^{7}+\tau^{8}) \big| {\widetilde S}_{5  2  1} \big>+\tau^{6} \big| {\widetilde S}_{5  1  1  1} \big> +(\tau^{4}+\tau^{6}+\tau^{8}) \big| {\widetilde S}_{4  4} \big>+(\tau^{5}+2 \tau^{6}+2 \tau^{7}+\tau^{8}+\tau^{9}) \big| {\widetilde S}_{4  3  1} \big>+(\tau^{6}+\tau^{7}+2 \tau^{8}+\tau^{9}+\tau^{10}) \big| {\widetilde S}_{4  2  2} \big> + (\tau^{7}+\tau^{8}+\tau^{9}) \big| {\widetilde S}_{4  2  1  1} \big>+(\tau^{7}+\tau^{8}+\tau^{9}) \big| {\widetilde S}_{3  3  2} \big>+(\tau^{8}+\tau^{10}) \big| {\widetilde S}_{3  3  1  1} \big>+(\tau^{9}+\tau^{10}+\tau^{11}) \big| {\widetilde S}_{3  2  2  1} \big>+\tau^{12} \big| {\widetilde S}_{2  2  2  2} \big>$
\end{center}

\begin{center}
$ \big| P^{(2,1)}_{[211]} \big> =
 \big| {\widetilde S}_{8} \big>+( q+\tau+\tau^{2}) \big| {\widetilde S}_{7  1} \big>+( q^2+\tau q+\tau^{2}+\tau^{2} q+\tau^{3}+\tau^{4}) \big| {\widetilde S}_{6  2} \big>+(\tau q+\tau^{2} q+\tau^{3}) \big| {\widetilde S}_{6  1  1} \big>+(\tau q^2+2 \tau^{2} q+\tau^{3}+\tau^{3} q+\tau^{4}) \big| {\widetilde S}_{5  3} \big>+(\tau q^2+\tau^{2} q+\tau^{2} q^2+2 \tau^{3} q+\tau^{4}+\tau^{4} q+\tau^{5}) \big| {\widetilde S}_{5  2  1} \big>+\tau^{3} q \big| {\widetilde S}_{5  1  1  1} \big> +(\tau^{2} q^2+\tau^{3} q+\tau^{4}) \big| {\widetilde S}_{4  4} \big>+(\tau^{2} q^2+\tau^{3} q^2+2 \tau^{3} q+2 \tau^{4} q+\tau^{5}) \big| {\widetilde S}_{4  3  1} \big>+(\tau^{2} q^2+\tau^{3} q^2+\tau^{4} q+\tau^{4} q^2+\tau^{5} q+\tau^{6}) \big| {\widetilde S}_{4  2  2} \big>+(\tau^{3} q^2+\tau^{4} q+\tau^{5} q) \big| {\widetilde S}_{4  2  1  1} \big>+(\tau^{3} q^2+\tau^{4} q+\tau^{5} q) \big| {\widetilde S}_{3  3  2} \big>+(\tau^{4} q^2+\tau^{5} q) \big| {\widetilde S}_{3  3  1  1} \big>+(\tau^{4} q^2+\tau^{5} q^2+\tau^{6} q) \big| {\widetilde S}_{3  2  2  1} \big>+\tau^{6} q^2 \big| {\widetilde S}_{2  2  2  2} \big>$
\end{center}

\begin{center}
$\big| P^{(2,1)}_{[31]} \big> =
 \big| {\widetilde S}_{8} \big>+( q+ q^2+\tau) \big| {\widetilde S}_{7  1} \big>+( q^4+ q^2+ q^3+\tau q+\tau q^2+\tau^{2}) \big| {\widetilde S}_{6  2} \big>+( q^3+\tau q+\tau q^2) \big| {\widetilde S}_{6  1  1} \big>+( q^3+ q^4+\tau q^3+2 \tau q^2+\tau^{2} q) \big| {\widetilde S}_{5  3} \big>+( q^4+ q^5+\tau q^2+2 \tau q^3+\tau q^4+\tau^{2} q+\tau^{2} q^2) \big| {\widetilde S}_{5  2  1} \big>+\tau q^3 \big| {\widetilde S}_{5  1  1  1} \big> +( q^4+\tau q^3+\tau^{2} q^2) \big| {\widetilde S}_{4  4} \big>+( q^5+2 \tau q^4+2 \tau q^3+\tau^{2} q^3+\tau^{2} q^2) \big| {\widetilde S}_{4  3  1} \big>+( q^6+\tau q^4+\tau q^5+\tau^{2} q^4+\tau^{2} q^2+\tau^{2} q^3) \big| {\widetilde S}_{4  2  2} \big>+(\tau q^4+\tau q^5+\tau^{2} q^3) \big| {\widetilde S}_{4  2  1  1} \big>+(\tau q^4+\tau q^5+\tau^{2} q^3) \big| {\widetilde S}_{3  3  2} \big>+(\tau q^5+\tau^{2} q^4) \big| {\widetilde S}_{3  3  1  1} \big>+(\tau q^6+\tau^{2} q^4+\tau^{2} q^5) \big| {\widetilde S}_{3  2  2  1} \big>+\tau^{2} q^6 \big| {\widetilde S}_{2  2  2  2} \big>$
\end{center}

\begin{center}
$ \big| P^{(2,1)}_{[4]} \big> =
 \big| {\widetilde S}_{8} \big>+( q^3+ q^2+ q) \big| {\widetilde S}_{7  1} \big>+( q^3+ q^5+ q^6+2  q^4+ q^2) \big| {\widetilde S}_{6  2} \big>+( q^5+ q^4+ q^3) \big| {\widetilde S}_{6  1  1} \big>+( q^7+2  q^5+ q^6+ q^4+ q^3) \big| {\widetilde S}_{5  3} \big>+(2 q^7+ q^8+2  q^6+2  q^5+ q^4) \big| {\widetilde S}_{5  2  1} \big>+ q^6 \big| {\widetilde S}_{5  1  1  1} \big> +( q^8+ q^4+ q^6) \big| {\widetilde S}_{4  4} \big>+( q^5+ q^9+2  q^7+ q^8+2  q^6) \big| {\widetilde S}_{4  3  1} \big>+( q^7+ q^9+ q^{10}+2 q^8+ q^6) \big| {\widetilde S}_{4  2  2} \big>+( q^7+ q^8+ q^9) \big| {\widetilde S}_{4  2  1  1} \big>+( q^7+ q^8+ q^9) \big| {\widetilde S}_{3  3  2} \big>+( q^{10} + q^8) \big| {\widetilde S}_{3  3  1  1} \big>+( q^9+ q^{10}+ q^{11}) \big| {\widetilde S}_{3  2  2  1} \big>+ q^{12} \big| {\widetilde S}_{2  2  2  2} \big> $
\end{center}

\begin{center}
$ \big| P^{(2,1)}_{[22]} \big> =
 \big| {\widetilde S}_{8} \big>+( q+\tau+\tau q) \big| {\widetilde S}_{7  1} \big>+( q^2+\tau q+\tau q^2+\tau^{2}+\tau^{2} q+\tau^{2} q^2) \big| {\widetilde S}_{6  2} \big>+(\tau q+\tau q^2+\tau^{2} q) \big| {\widetilde S}_{6  1  1} \big>+( q^3+\tau q^3+\tau q^2+\tau^{2} q+\tau^{3}+\tau^{3} q) \big| {\widetilde S}_{5  3} \big>+(\tau q^3+\tau q^2+2 \tau^{2} q^2+\tau^{2} q+\tau^{2} q^3+\tau^{3} q+\tau^{3} q^2) \big| {\widetilde S}_{5  2  1} \big>+\tau^{2} q^2 \big| {\widetilde S}_{5  1  1  1} \big> +( q^4+\tau^{2} q^2+\tau^{4}) \big| {\widetilde S}_{4  4} \big>+(\tau q^3+\tau q^4+\tau^{2} q^3+\tau^{2} q^2+\tau^{3} q+\tau^{3} q^2+\tau^{4} q) \big| {\widetilde S}_{4  3  1} \big>+(\tau^{2} q^4+\tau^{2} q^2+\tau^{2} q^3+\tau^{3} q^3+\tau^{3} q^2+\tau^{4} q^2) \big| {\widetilde S}_{4  2  2} \big>+(\tau^{2} q^3+\tau^{3} q^3+\tau^{3} q^2) \big| {\widetilde S}_{4  2  1  1} \big>+(\tau^{2} q^3+\tau^{3} q^3+\tau^{3} q^2) \big| {\widetilde S}_{3  3  2} \big>+(\tau^{2} q^4+\tau^{4} q^2) \big| {\widetilde S}_{3  3  1  1} \big>+(\tau^{3} q^3+\tau^{3} q^4+\tau^{4} q^3) \big| {\widetilde S}_{3  2  2  1} \big>+\tau^{4} q^4 \big| {\widetilde S}_{2  2  2  2} \big> $
\end{center}

\subsubsection*{Representations of size $|R| = 5$}

{\fontsize{8pt}{0pt}

\begin{center}
$ \big| P^{(2,1)}_{[11111]} \big> =
\big| {\widetilde S}_{10} \big>+(\tau+\tau^{2}+\tau^{3}+\tau^{4}) \big| {\widetilde S}_{9  1} \big>+(\tau^{2}+\tau^{3}+2 \tau^{4}+2 \tau^{5}+2 \tau^{6}+\tau^{7}+\tau^{8}) \big| {\widetilde S}_{8  2} \big>+(\tau^{3}+\tau^{4}+2 \tau^{5}+\tau^{6}+\tau^{7}) \big| {\widetilde S}_{8  1  1} \big>+(\tau^{3}+\tau^{4}+2 \tau^{5}+3 \tau^{6}+3 \tau^{7}+2 \tau^{8}+2 \tau^{9}+\tau^{10}) \big| {\widetilde S}_{7  3} \big>+(\tau^{4}+2 \tau^{5}+3 \tau^{6}+4 \tau^{7}+4 \tau^{8}+3 \tau^{9}+2 \tau^{10}+\tau^{11}) \big| {\widetilde S}_{7  2  1} \big>+(\tau^{6}+\tau^{7}+\tau^{8}+\tau^{9}) \big| {\widetilde S}_{7  1  1  1} \big>+(\tau^{4}+\tau^{5}+2 \tau^{6}+2 \tau^{7}+3 \tau^{8}+2 \tau^{9}+2 \tau^{10}+\tau^{11}+\tau^{12}) \big| {\widetilde S}_{6  4} \big>+(\tau^{5}+2 \tau^{6}+4 \tau^{7}+5 \tau^{8}+6 \tau^{9}+5 \tau^{10}+4 \tau^{11}+2 \tau^{12}+\tau^{13}) \big| {\widetilde S}_{6  3  1} \big>+(\tau^{6}+\tau^{7}+3 \tau^{8}+3 \tau^{9}+4 \tau^{10}+3 \tau^{11}+3 \tau^{12}+\tau^{13}+\tau^{14}) \big| {\widetilde S}_{6  2  2} \big>+(\tau^{7}+2 \tau^{8}+3 \tau^{9}+3 \tau^{10}+3 \tau^{11}+2 \tau^{12}+\tau^{13}) \big| {\widetilde S}_{6  2  1  1} \big>+\tau^{10} \big| {\widetilde S}_{6  1  1  1  1} \big>+(\tau^{5}+\tau^{7}+\tau^{8}+\tau^{9}+\tau^{10}+\tau^{11}) \big| {\widetilde S}_{5  5} \big>+(\tau^{6}+2 \tau^{7}+3 \tau^{8}+4 \tau^{9}+4 \tau^{10}+4 \tau^{11}+3 \tau^{12}+2 \tau^{13}+\tau^{14}) \big| {\widetilde S}_{5  4  1} \big>+(\tau^{7}+2 \tau^{8}+4 \tau^{9}+5 \tau^{10}+6 \tau^{11}+5 \tau^{12}+4 \tau^{13}+2 \tau^{14}+\tau^{15}) \big| {\widetilde S}_{5  3  2} \big>+(\tau^{8}+2 \tau^{9}+4 \tau^{10}+4 \tau^{11}+4 \tau^{12}+3 \tau^{13}+2 \tau^{14}+\tau^{15}) \big| {\widetilde S}_{5  3  1  1} \big>+(\tau^{9}+2 \tau^{10}+3 \tau^{11}+4 \tau^{12}+4 \tau^{13}+3 \tau^{14}+2 \tau^{15}+\tau^{16}) \big| {\widetilde S}_{5  2  2  1} \big>+(\tau^{11}+\tau^{12}+\tau^{13}+\tau^{14}) \big| {\widetilde S}_{5  2  1  1  1} \big>+(\tau^{8}+\tau^{9}+3 \tau^{10}+2 \tau^{11}+3 \tau^{12}+2 \tau^{13}+2 \tau^{14}+\tau^{15}+\tau^{16}) \big| {\widetilde S}_{4  4  2} \big>+(\tau^{9}+\tau^{10}+2 \tau^{11}+2 \tau^{12}+2 \tau^{13}+\tau^{14}+\tau^{15}) \big| {\widetilde S}_{4  4  1  1} \big>+(\tau^{9}+\tau^{10}+2 \tau^{11}+2 \tau^{12}+2 \tau^{13}+\tau^{14}+\tau^{15}) \big| {\widetilde S}_{4  3  3} \big>+(\tau^{10}+3 \tau^{11}+4 \tau^{12}+5 \tau^{13}+5 \tau^{14}+3 \tau^{15}+2 \tau^{16}+\tau^{17}) \big| {\widetilde S}_{4  3  2  1} \big>+(\tau^{12}+\tau^{13}+\tau^{14}+\tau^{15}+\tau^{16}) \big| {\widetilde S}_{4  3  1  1  1} \big>+(\tau^{12}+\tau^{13}+2 \tau^{14}+2 \tau^{15}+2 \tau^{16}+\tau^{17}+\tau^{18}) \big| {\widetilde S}_{4  2  2  2} \big>+(\tau^{13}+\tau^{14}+2 \tau^{15}+\tau^{16}+\tau^{17}) \big| {\widetilde S}_{4  2  2  1  1} \big>+(\tau^{12}+\tau^{13}+\tau^{14}+\tau^{15}+\tau^{16}) \big| {\widetilde S}_{3  3  3  1} \big>+(\tau^{13}+\tau^{14}+2 \tau^{15}+\tau^{16}+\tau^{17}) \big| {\widetilde S}_{3  3  2  2} \big>+(\tau^{14}+\tau^{15}+\tau^{16}+\tau^{17}+\tau^{18}) \big| {\widetilde S}_{3  3  2  1  1} \big>+(\tau^{16}+\tau^{17}+\tau^{18}+\tau^{19}) \big| {\widetilde S}_{3  2  2  2  1} \big>+\tau^{20} \big| {\widetilde S}_{2  2  2  2  2} \big>
$
\end{center}

\begin{center}
$ \big| P^{(2,1)}_{[2111]} \big> =
\big| {\widetilde S}_{10} \big>+(\tau+\tau^{2}+\tau^{3}+q ) \big| {\widetilde S}_{9  1} \big>+(\tau^{2}+\tau^{3}+2 \tau^{4}+\tau^{5}+\tau^{6}+\tau q +\tau^{2} q +\tau^{3} q +q^{2}) \big| {\widetilde S}_{8  2} \big>+(\tau^{3}+\tau^{4}+\tau^{5}+\tau q +\tau^{2} q +\tau^{3} q ) \big| {\widetilde S}_{8  1  1} \big>+(\tau^{3}+\tau^{4}+2 \tau^{5}+2 \tau^{6}+\tau^{7}+\tau^{2} q +2 \tau^{3} q +2 \tau^{4} q +\tau^{5} q +\tau q^{2}+\tau^{2} q^{2}) \big| {\widetilde S}_{7  3} \big>+(\tau^{4}+2 \tau^{5}+2 \tau^{6}+2 \tau^{7}+\tau^{8}+\tau^{2} q +2 \tau^{3} q +3 \tau^{4} q +2 \tau^{5} q +\tau^{6} q +\tau q^{2}+\tau^{2} q^{2}+\tau^{3} q^{2}) \big| {\widetilde S}_{7  2  1} \big>+(\tau^{6}+\tau^{3} q +\tau^{4} q +\tau^{5} q ) \big| {\widetilde S}_{7  1  1  1} \big>+(\tau^{4}+\tau^{5}+2 \tau^{6}+\tau^{7}+\tau^{8}+\tau^{3} q +2 \tau^{4} q +2 \tau^{5} q +\tau^{6} q +\tau^{2} q^{2}+\tau^{3} q^{2}+\tau^{4} q^{2}) \big| {\widetilde S}_{6  4} \big>+(\tau^{5}+2 \tau^{6}+3 \tau^{7}+2 \tau^{8}+\tau^{9}+\tau^{3} q +3 \tau^{4} q +5 \tau^{5} q +4 \tau^{6} q +2 \tau^{7} q +\tau^{2} q^{2}+2 \tau^{3} q^{2}+2 \tau^{4} q^{2}+\tau^{5} q^{2}) \big| {\widetilde S}_{6  3  1} \big>+(\tau^{6}+\tau^{7}+2 \tau^{8}+\tau^{9}+\tau^{10}+\tau^{4} q +2 \tau^{5} q +2 \tau^{6} q +2 \tau^{7} q +\tau^{8} q +\tau^{2} q^{2}+\tau^{3} q^{2}+2 \tau^{4} q^{2}+\tau^{5} q^{2}+\tau^{6} q^{2}) \big| {\widetilde S}_{6  2  2} \big>+(\tau^{7}+\tau^{8}+\tau^{9}+\tau^{4} q +2 \tau^{5} q +3 \tau^{6} q +2 \tau^{7} q +\tau^{8} q +\tau^{3} q^{2}+\tau^{4} q^{2}+\tau^{5} q^{2}) \big| {\widetilde S}_{6  2  1  1} \big>+\tau^{6} q  \big| {\widetilde S}_{6  1  1  1  1} \big>+(\tau^{5}+\tau^{7}+\tau^{4} q +\tau^{5} q +\tau^{6} q +\tau^{3} q^{2}) \big| {\widetilde S}_{5  5} \big>+(\tau^{6}+2 \tau^{7}+2 \tau^{8}+\tau^{9}+\tau^{4} q +3 \tau^{5} q +4 \tau^{6} q +3 \tau^{7} q +\tau^{8} q +\tau^{3} q^{2}+2 \tau^{4} q^{2}+2 \tau^{5} q^{2}+\tau^{6} q^{2}) \big| {\widetilde S}_{5  4  1} \big>+(\tau^{7}+2 \tau^{8}+2 \tau^{9}+\tau^{10}+2 \tau^{5} q +4 \tau^{6} q +5 \tau^{7} q +3 \tau^{8} q +\tau^{9} q +\tau^{3} q^{2}+2 \tau^{4} q^{2}+3 \tau^{5} q^{2}+2 \tau^{6} q^{2}+\tau^{7} q^{2}) \big| {\widetilde S}_{5  3  2} \big>+(\tau^{8}+\tau^{9}+\tau^{10}+\tau^{5} q +3 \tau^{6} q +4 \tau^{7} q +3 \tau^{8} q +\tau^{9} q +\tau^{4} q^{2}+2 \tau^{5} q^{2}+2 \tau^{6} q^{2}+\tau^{7} q^{2}) \big| {\widetilde S}_{5  3  1  1} \big>+(\tau^{9}+\tau^{10}+\tau^{11}+\tau^{6} q +2 \tau^{7} q +3 \tau^{8} q +2 \tau^{9} q +\tau^{10} q +\tau^{4} q^{2}+2 \tau^{5} q^{2}+2 \tau^{6} q^{2}+2 \tau^{7} q^{2}+\tau^{8} q^{2}) \big| {\widetilde S}_{5  2  2  1} \big>+(\tau^{7} q +\tau^{8} q +\tau^{9} q +\tau^{6} q^{2}) \big| {\widetilde S}_{5  2  1  1  1} \big>+(\tau^{8}+\tau^{9}+\tau^{10}+2 \tau^{6} q +2 \tau^{7} q +2 \tau^{8} q +\tau^{9} q +\tau^{4} q^{2}+\tau^{5} q^{2}+2 \tau^{6} q^{2}+\tau^{7} q^{2}+\tau^{8} q^{2}) \big| {\widetilde S}_{4  4  2} \big>+(\tau^{9}+\tau^{6} q +2 \tau^{7} q +2 \tau^{8} q +\tau^{9} q +\tau^{5} q^{2}+\tau^{6} q^{2}+\tau^{7} q^{2}) \big| {\widetilde S}_{4  4  1  1} \big>+(\tau^{9}+\tau^{6} q +2 \tau^{7} q +2 \tau^{8} q +\tau^{9} q +\tau^{5} q^{2}+\tau^{6} q^{2}+\tau^{7} q^{2}) \big| {\widetilde S}_{4  3  3} \big>+(\tau^{10}+\tau^{11}+2 \tau^{7} q +4 \tau^{8} q +4 \tau^{9} q +2 \tau^{10} q +\tau^{5} q^{2}+3 \tau^{6} q^{2}+3 \tau^{7} q^{2}+2 \tau^{8} q^{2}+\tau^{9} q^{2}) \big| {\widetilde S}_{4  3  2  1} \big>+(\tau^{8} q +\tau^{9} q +\tau^{10} q +\tau^{7} q^{2}+\tau^{8} q^{2}) \big| {\widetilde S}_{4  3  1  1  1} \big>+(\tau^{12}+\tau^{9} q +\tau^{10} q +\tau^{11} q +\tau^{6} q^{2}+\tau^{7} q^{2}+2 \tau^{8} q^{2}+\tau^{9} q^{2}+\tau^{10} q^{2}) \big| {\widetilde S}_{4  2  2  2} \big>+(\tau^{9} q +\tau^{10} q +\tau^{11} q +\tau^{7} q^{2}+\tau^{8} q^{2}+\tau^{9} q^{2}) \big| {\widetilde S}_{4  2  2  1  1} \big>+(\tau^{8} q +\tau^{9} q +\tau^{10} q +\tau^{7} q^{2}+\tau^{8} q^{2}) \big| {\widetilde S}_{3  3  3  1} \big>+(\tau^{9} q +\tau^{10} q +\tau^{11} q +\tau^{7} q^{2}+\tau^{8} q^{2}+\tau^{9} q^{2}) \big| {\widetilde S}_{3  3  2  2} \big>+(\tau^{10} q +\tau^{11} q +\tau^{8} q^{2}+\tau^{9} q^{2}+\tau^{10} q^{2}) \big| {\widetilde S}_{3  3  2  1  1} \big>+(\tau^{12} q +\tau^{9} q^{2}+\tau^{10} q^{2}+\tau^{11} q^{2}) \big| {\widetilde S}_{3  2  2  2  1} \big>+\tau^{12} q^{2} \big| {\widetilde S}_{2  2  2  2  2} \big>
$
\end{center}

\begin{center}
$ \big| P^{(2,1)}_{[221]} \big> =
\big| {\widetilde S}_{10} \big>+(\tau+\tau^{2}+q +\tau q ) \big| {\widetilde S}_{9  1} \big>+(\tau^{2}+\tau^{3}+\tau^{4}+\tau q +2 \tau^{2} q +\tau^{3} q +q^{2}+\tau q^{2}+\tau^{2} q^{2}) \big| {\widetilde S}_{8  2} \big>+(\tau^{3}+\tau q +2 \tau^{2} q +\tau^{3} q +\tau q^{2}) \big| {\widetilde S}_{8  1  1} \big>+(\tau^{3}+\tau^{4}+\tau^{5}+\tau^{2} q +3 \tau^{3} q +2 \tau^{4} q +\tau q^{2}+2 \tau^{2} q^{2}+\tau^{3} q^{2}+q^{3}+\tau q^{3}) \big| {\widetilde S}_{7  3} \big>+(\tau^{4}+\tau^{5}+\tau^{2} q +3 \tau^{3} q +3 \tau^{4} q +\tau^{5} q +\tau q^{2}+3 \tau^{2} q^{2}+3 \tau^{3} q^{2}+\tau^{4} q^{2}+\tau q^{3}+\tau^{2} q^{3}) \big| {\widetilde S}_{7  2  1} \big>+(\tau^{3} q +\tau^{4} q +\tau^{2} q^{2}+\tau^{3} q^{2}) \big| {\widetilde S}_{7  1  1  1} \big>+(\tau^{4}+\tau^{5}+\tau^{6}+\tau^{3} q +2 \tau^{4} q +\tau^{5} q +2 \tau^{2} q^{2}+2 \tau^{3} q^{2}+\tau^{4} q^{2}+\tau q^{3}+\tau^{2} q^{3}+q^{4}) \big| {\widetilde S}_{6  4} \big>+(\tau^{5}+\tau^{6}+\tau^{3} q +4 \tau^{4} q +4 \tau^{5} q +\tau^{6} q +\tau^{2} q^{2}+5 \tau^{3} q^{2}+4 \tau^{4} q^{2}+\tau^{5} q^{2}+\tau q^{3}+3 \tau^{2} q^{3}+2 \tau^{3} q^{3}+\tau q^{4}) \big| {\widetilde S}_{6  3  1} \big>+(\tau^{6}+\tau^{4} q +2 \tau^{5} q +\tau^{6} q +\tau^{2} q^{2}+2 \tau^{3} q^{2}+4 \tau^{4} q^{2}+2 \tau^{5} q^{2}+\tau^{6} q^{2}+\tau^{2} q^{3}+2 \tau^{3} q^{3}+\tau^{4} q^{3}+\tau^{2} q^{4}) \big| {\widetilde S}_{6  2  2} \big>+(\tau^{4} q +2 \tau^{5} q +\tau^{6} q +2 \tau^{3} q^{2}+3 \tau^{4} q^{2}+2 \tau^{5} q^{2}+\tau^{2} q^{3}+2 \tau^{3} q^{3}+\tau^{4} q^{3}) \big| {\widetilde S}_{6  2  1  1} \big>+\tau^{4} q^{2} \big| {\widetilde S}_{6  1  1  1  1} \big>+(\tau^{5}+\tau^{4} q +\tau^{5} q +\tau^{3} q^{2}+\tau q^{3}+\tau^{2} q^{3}) \big| {\widetilde S}_{5  5} \big>+(\tau^{6}+\tau^{7}+\tau^{4} q +3 \tau^{5} q +2 \tau^{6} q +2 \tau^{3} q^{2}+4 \tau^{4} q^{2}+2 \tau^{5} q^{2}+2 \tau^{2} q^{3}+3 \tau^{3} q^{3}+\tau^{4} q^{3}+\tau q^{4}+\tau^{2} q^{4}) \big| {\widetilde S}_{5  4  1} \big>+(\tau^{7}+2 \tau^{5} q +3 \tau^{6} q +\tau^{7} q +\tau^{3} q^{2}+4 \tau^{4} q^{2}+5 \tau^{5} q^{2}+\tau^{6} q^{2}+\tau^{2} q^{3}+4 \tau^{3} q^{3}+4 \tau^{4} q^{3}+\tau^{5} q^{3}+\tau^{2} q^{4}+\tau^{3} q^{4}) \big| {\widetilde S}_{5  3  2} \big>+(\tau^{5} q +2 \tau^{6} q +\tau^{7} q +3 \tau^{4} q^{2}+4 \tau^{5} q^{2}+2 \tau^{6} q^{2}+2 \tau^{3} q^{3}+3 \tau^{4} q^{3}+\tau^{5} q^{3}+\tau^{2} q^{4}+\tau^{3} q^{4}) \big| {\widetilde S}_{5  3  1  1} \big>+(\tau^{6} q +\tau^{7} q +\tau^{4} q^{2}+3 \tau^{5} q^{2}+3 \tau^{6} q^{2}+\tau^{7} q^{2}+\tau^{3} q^{3}+3 \tau^{4} q^{3}+3 \tau^{5} q^{3}+\tau^{6} q^{3}+\tau^{3} q^{4}+\tau^{4} q^{4}) \big| {\widetilde S}_{5  2  2  1} \big>+(\tau^{5} q^{2}+\tau^{6} q^{2}+\tau^{4} q^{3}+\tau^{5} q^{3}) \big| {\widetilde S}_{5  2  1  1  1} \big>+(\tau^{8}+\tau^{6} q +\tau^{7} q +2 \tau^{4} q^{2}+2 \tau^{5} q^{2}+2 \tau^{6} q^{2}+\tau^{3} q^{3}+2 \tau^{4} q^{3}+\tau^{5} q^{3}+\tau^{2} q^{4}+\tau^{3} q^{4}+\tau^{4} q^{4}) \big| {\widetilde S}_{4  4  2} \big>+(\tau^{6} q +\tau^{7} q +2 \tau^{5} q^{2}+\tau^{6} q^{2}+\tau^{3} q^{3}+2 \tau^{4} q^{3}+\tau^{5} q^{3}+\tau^{3} q^{4}) \big| {\widetilde S}_{4  4  1  1} \big>+(\tau^{6} q +\tau^{7} q +2 \tau^{5} q^{2}+\tau^{6} q^{2}+\tau^{3} q^{3}+2 \tau^{4} q^{3}+\tau^{5} q^{3}+\tau^{3} q^{4}) \big| {\widetilde S}_{4  3  3} \big>+(\tau^{7} q +\tau^{8} q +2 \tau^{5} q^{2}+4 \tau^{6} q^{2}+2 \tau^{7} q^{2}+3 \tau^{4} q^{3}+5 \tau^{5} q^{3}+2 \tau^{6} q^{3}+\tau^{3} q^{4}+2 \tau^{4} q^{4}+\tau^{5} q^{4}) \big| {\widetilde S}_{4  3  2  1} \big>+(\tau^{6} q^{2}+\tau^{7} q^{2}+\tau^{5} q^{3}+\tau^{6} q^{3}+\tau^{4} q^{4}) \big| {\widetilde S}_{4  3  1  1  1} \big>+(\tau^{6} q^{2}+\tau^{7} q^{2}+\tau^{8} q^{2}+\tau^{5} q^{3}+2 \tau^{6} q^{3}+\tau^{7} q^{3}+\tau^{4} q^{4}+\tau^{5} q^{4}+\tau^{6} q^{4}) \big| {\widetilde S}_{4  2  2  2} \big>+(\tau^{7} q^{2}+\tau^{5} q^{3}+2 \tau^{6} q^{3}+\tau^{7} q^{3}+\tau^{5} q^{4}) \big| {\widetilde S}_{4  2  2  1  1} \big>+(\tau^{6} q^{2}+\tau^{7} q^{2}+\tau^{5} q^{3}+\tau^{6} q^{3}+\tau^{4} q^{4}) \big| {\widetilde S}_{3  3  3  1} \big>+(\tau^{7} q^{2}+\tau^{5} q^{3}+2 \tau^{6} q^{3}+\tau^{7} q^{3}+\tau^{5} q^{4}) \big| {\widetilde S}_{3  3  2  2} \big>+(\tau^{8} q^{2}+\tau^{6} q^{3}+\tau^{7} q^{3}+\tau^{5} q^{4}+\tau^{6} q^{4}) \big| {\widetilde S}_{3  3  2  1  1} \big>+(\tau^{7} q^{3}+\tau^{8} q^{3}+\tau^{6} q^{4}+\tau^{7} q^{4}) \big| {\widetilde S}_{3  2  2  2  1} \big>+\tau^{8} q^{4} \big| {\widetilde S}_{2  2  2  2  2} \big>
$
\end{center}

\begin{center}
$ \big| P^{(2,1)}_{[311]} \big> =
\big| {\widetilde S}_{10} \big>+(\tau+\tau^{2}+q +q^{2}) \big| {\widetilde S}_{9  1} \big>+(\tau^{2}+\tau^{3}+\tau^{4}+\tau q +\tau^{2} q +q^{2}+\tau q^{2}+\tau^{2} q^{2}+q^{3}+q^{4}) \big| {\widetilde S}_{8  2} \big>+(\tau^{3}+\tau q +\tau^{2} q +\tau q^{2}+\tau^{2} q^{2}+q^{3}) \big| {\widetilde S}_{8  1  1} \big>+(\tau^{3}+\tau^{4}+\tau^{2} q +\tau^{3} q +\tau^{4} q +\tau q^{2}+3 \tau^{2} q^{2}+\tau^{3} q^{2}+q^{3}+\tau q^{3}+\tau^{2} q^{3}+q^{4}+\tau q^{4}) \big| {\widetilde S}_{7  3} \big>+(\tau^{4}+\tau^{5}+\tau^{2} q +2 \tau^{3} q +\tau^{4} q +\tau q^{2}+2 \tau^{2} q^{2}+2 \tau^{3} q^{2}+\tau^{4} q^{2}+2 \tau q^{3}+2 \tau^{2} q^{3}+q^{4}+\tau q^{4}+\tau^{2} q^{4}+q^{5}) \big| {\widetilde S}_{7  2  1} \big>+(\tau^{3} q +\tau^{3} q^{2}+\tau q^{3}+\tau^{2} q^{3}) \big| {\widetilde S}_{7  1  1  1} \big>+(\tau^{4}+\tau^{3} q +\tau^{4} q +2 \tau^{2} q^{2}+2 \tau^{3} q^{2}+\tau^{4} q^{2}+\tau q^{3}+2 \tau^{2} q^{3}+\tau^{3} q^{3}+q^{4}+\tau q^{4}+\tau^{2} q^{4}) \big| {\widetilde S}_{6  4} \big>+(\tau^{5}+\tau^{3} q +2 \tau^{4} q +\tau^{5} q +\tau^{2} q^{2}+4 \tau^{3} q^{2}+3 \tau^{4} q^{2}+\tau q^{3}+4 \tau^{2} q^{3}+3 \tau^{3} q^{3}+\tau^{4} q^{3}+2 \tau q^{4}+3 \tau^{2} q^{4}+\tau^{3} q^{4}+q^{5}+\tau q^{5}) \big| {\widetilde S}_{6  3  1} \big>+(\tau^{6}+\tau^{4} q +\tau^{5} q +\tau^{2} q^{2}+\tau^{3} q^{2}+2 \tau^{4} q^{2}+\tau^{5} q^{2}+\tau^{2} q^{3}+2 \tau^{3} q^{3}+\tau^{4} q^{3}+\tau q^{4}+2 \tau^{2} q^{4}+\tau^{3} q^{4}+\tau^{4} q^{4}+\tau q^{5}+\tau^{2} q^{5}+q^{6}) \big| {\widetilde S}_{6  2  2} \big>+(\tau^{4} q +\tau^{5} q +\tau^{3} q^{2}+\tau^{4} q^{2}+\tau^{5} q^{2}+\tau^{2} q^{3}+3 \tau^{3} q^{3}+\tau^{4} q^{3}+\tau q^{4}+\tau^{2} q^{4}+\tau^{3} q^{4}+\tau q^{5}+\tau^{2} q^{5}) \big| {\widetilde S}_{6  2  1  1} \big>+\tau^{3} q^{3} \big| {\widetilde S}_{6  1  1  1  1} \big>+(\tau^{4} q +\tau^{2} q^{2}+\tau^{3} q^{2}+\tau^{2} q^{3}+\tau^{3} q^{3}+\tau q^{4}) \big| {\widetilde S}_{5  5} \big>+(\tau^{4} q +\tau^{5} q +2 \tau^{3} q^{2}+3 \tau^{4} q^{2}+\tau^{5} q^{2}+2 \tau^{2} q^{3}+4 \tau^{3} q^{3}+2 \tau^{4} q^{3}+\tau q^{4}+3 \tau^{2} q^{4}+2 \tau^{3} q^{4}+\tau q^{5}+\tau^{2} q^{5}) \big| {\widetilde S}_{5  4  1} \big>+(\tau^{5} q +\tau^{6} q +\tau^{3} q^{2}+3 \tau^{4} q^{2}+2 \tau^{5} q^{2}+\tau^{2} q^{3}+3 \tau^{3} q^{3}+4 \tau^{4} q^{3}+\tau^{5} q^{3}+3 \tau^{2} q^{4}+4 \tau^{3} q^{4}+\tau^{4} q^{4}+\tau q^{5}+2 \tau^{2} q^{5}+\tau^{3} q^{5}+\tau q^{6}) \big| {\widetilde S}_{5  3  2} \big>+(\tau^{5} q +\tau^{4} q^{2}+2 \tau^{5} q^{2}+3 \tau^{3} q^{3}+3 \tau^{4} q^{3}+\tau^{5} q^{3}+\tau^{2} q^{4}+3 \tau^{3} q^{4}+2 \tau^{4} q^{4}+\tau q^{5}+2 \tau^{2} q^{5}+\tau^{3} q^{5}) \big| {\widetilde S}_{5  3  1  1} \big>+(\tau^{6} q +\tau^{4} q^{2}+\tau^{5} q^{2}+\tau^{6} q^{2}+2 \tau^{4} q^{3}+2 \tau^{5} q^{3}+\tau^{2} q^{4}+2 \tau^{3} q^{4}+2 \tau^{4} q^{4}+\tau^{5} q^{4}+\tau^{2} q^{5}+2 \tau^{3} q^{5}+\tau^{4} q^{5}+\tau q^{6}+\tau^{2} q^{6}) \big| {\widetilde S}_{5  2  2  1} \big>+(\tau^{4} q^{3}+\tau^{5} q^{3}+\tau^{3} q^{4}+\tau^{3} q^{5}) \big| {\widetilde S}_{5  2  1  1  1} \big>+(\tau^{4} q^{2}+\tau^{5} q^{2}+\tau^{6} q^{2}+2 \tau^{3} q^{3}+2 \tau^{4} q^{3}+\tau^{5} q^{3}+\tau^{2} q^{4}+2 \tau^{3} q^{4}+2 \tau^{4} q^{4}+\tau^{2} q^{5}+\tau^{3} q^{5}+\tau^{2} q^{6}) \big| {\widetilde S}_{4  4  2} \big>+(\tau^{5} q^{2}+\tau^{3} q^{3}+2 \tau^{4} q^{3}+\tau^{5} q^{3}+2 \tau^{3} q^{4}+\tau^{4} q^{4}+\tau^{2} q^{5}+\tau^{3} q^{5}) \big| {\widetilde S}_{4  4  1  1} \big>+(\tau^{5} q^{2}+\tau^{3} q^{3}+2 \tau^{4} q^{3}+\tau^{5} q^{3}+2 \tau^{3} q^{4}+\tau^{4} q^{4}+\tau^{2} q^{5}+\tau^{3} q^{5}) \big| {\widetilde S}_{4  3  3} \big>+(\tau^{5} q^{2}+\tau^{6} q^{2}+2 \tau^{4} q^{3}+3 \tau^{5} q^{3}+\tau^{6} q^{3}+2 \tau^{3} q^{4}+4 \tau^{4} q^{4}+2 \tau^{5} q^{4}+\tau^{2} q^{5}+3 \tau^{3} q^{5}+2 \tau^{4} q^{5}+\tau^{2} q^{6}+\tau^{3} q^{6}) \big| {\widetilde S}_{4  3  2  1} \big>+(\tau^{5} q^{3}+\tau^{4} q^{4}+\tau^{5} q^{4}+\tau^{3} q^{5}+\tau^{4} q^{5}) \big| {\widetilde S}_{4  3  1  1  1} \big>+(\tau^{6} q^{2}+\tau^{6} q^{3}+\tau^{4} q^{4}+\tau^{5} q^{4}+\tau^{6} q^{4}+\tau^{4} q^{5}+\tau^{5} q^{5}+\tau^{2} q^{6}+\tau^{3} q^{6}+\tau^{4} q^{6}) \big| {\widetilde S}_{4  2  2  2} \big>+(\tau^{6} q^{3}+\tau^{4} q^{4}+\tau^{5} q^{4}+\tau^{4} q^{5}+\tau^{5} q^{5}+\tau^{3} q^{6}) \big| {\widetilde S}_{4  2  2  1  1} \big>+(\tau^{5} q^{3}+\tau^{4} q^{4}+\tau^{5} q^{4}+\tau^{3} q^{5}+\tau^{4} q^{5}) \big| {\widetilde S}_{3  3  3  1} \big>+(\tau^{6} q^{3}+\tau^{4} q^{4}+\tau^{5} q^{4}+\tau^{4} q^{5}+\tau^{5} q^{5}+\tau^{3} q^{6}) \big| {\widetilde S}_{3  3  2  2} \big>+(\tau^{5} q^{4}+\tau^{6} q^{4}+\tau^{4} q^{5}+\tau^{5} q^{5}+\tau^{4} q^{6}) \big| {\widetilde S}_{3  3  2  1  1} \big>+(\tau^{6} q^{4}+\tau^{6} q^{5}+\tau^{4} q^{6}+\tau^{5} q^{6}) \big| {\widetilde S}_{3  2  2  2  1} \big>+\tau^{6} q^{6} \big| {\widetilde S}_{2  2  2  2  2} \big>
$
\end{center}

\begin{center}
$ \big| P^{(2,1)}_{[32]} \big> =
\big| {\widetilde S}_{10} \big>+(\tau+q +\tau q +q^{2}) \big| {\widetilde S}_{9  1} \big>+(\tau^{2}+\tau q +\tau^{2} q +q^{2}+2 \tau q^{2}+\tau^{2} q^{2}+q^{3}+\tau q^{3}+q^{4}) \big| {\widetilde S}_{8  2} \big>+(\tau q +\tau^{2} q +2 \tau q^{2}+q^{3}+\tau q^{3}) \big| {\widetilde S}_{8  1  1} \big>+(\tau^{3}+\tau^{2} q +\tau^{3} q +\tau q^{2}+2 \tau^{2} q^{2}+q^{3}+3 \tau q^{3}+\tau^{2} q^{3}+q^{4}+2 \tau q^{4}+q^{5}) \big| {\widetilde S}_{7  3} \big>+(\tau^{2} q +\tau^{3} q +\tau q^{2}+3 \tau^{2} q^{2}+\tau^{3} q^{2}+3 \tau q^{3}+3 \tau^{2} q^{3}+q^{4}+3 \tau q^{4}+\tau^{2} q^{4}+q^{5}+\tau q^{5}) \big| {\widetilde S}_{7  2  1} \big>+(\tau^{2} q^{2}+\tau q^{3}+\tau^{2} q^{3}+\tau q^{4}) \big| {\widetilde S}_{7  1  1  1} \big>+(\tau^{4}+\tau^{3} q +2 \tau^{2} q^{2}+\tau^{3} q^{2}+\tau q^{3}+2 \tau^{2} q^{3}+q^{4}+2 \tau q^{4}+\tau^{2} q^{4}+q^{5}+\tau q^{5}+q^{6}) \big| {\widetilde S}_{6  4} \big>+(\tau^{3} q +\tau^{4} q +\tau^{2} q^{2}+3 \tau^{3} q^{2}+\tau q^{3}+5 \tau^{2} q^{3}+2 \tau^{3} q^{3}+4 \tau q^{4}+4 \tau^{2} q^{4}+q^{5}+4 \tau q^{5}+\tau^{2} q^{5}+q^{6}+\tau q^{6}) \big| {\widetilde S}_{6  3  1} \big>+(\tau^{2} q^{2}+\tau^{3} q^{2}+\tau^{4} q^{2}+2 \tau^{2} q^{3}+2 \tau^{3} q^{3}+\tau q^{4}+4 \tau^{2} q^{4}+\tau^{3} q^{4}+2 \tau q^{5}+2 \tau^{2} q^{5}+q^{6}+\tau q^{6}+\tau^{2} q^{6}) \big| {\widetilde S}_{6  2  2} \big>+(\tau^{3} q^{2}+2 \tau^{2} q^{3}+2 \tau^{3} q^{3}+\tau q^{4}+3 \tau^{2} q^{4}+\tau^{3} q^{4}+2 \tau q^{5}+2 \tau^{2} q^{5}+\tau q^{6}) \big| {\widetilde S}_{6  2  1  1} \big>+\tau^{2} q^{4} \big| {\widetilde S}_{6  1  1  1  1} \big>+(\tau^{3} q +\tau^{3} q^{2}+\tau^{2} q^{3}+\tau q^{4}+q^{5}+\tau q^{5}) \big| {\widetilde S}_{5  5} \big>+(\tau^{4} q +2 \tau^{3} q^{2}+\tau^{4} q^{2}+2 \tau^{2} q^{3}+3 \tau^{3} q^{3}+\tau q^{4}+4 \tau^{2} q^{4}+\tau^{3} q^{4}+3 \tau q^{5}+2 \tau^{2} q^{5}+q^{6}+2 \tau q^{6}+q^{7}) \big| {\widetilde S}_{5  4  1} \big>+(\tau^{3} q^{2}+\tau^{4} q^{2}+\tau^{2} q^{3}+4 \tau^{3} q^{3}+\tau^{4} q^{3}+4 \tau^{2} q^{4}+4 \tau^{3} q^{4}+2 \tau q^{5}+5 \tau^{2} q^{5}+\tau^{3} q^{5}+3 \tau q^{6}+\tau^{2} q^{6}+q^{7}+\tau q^{7}) \big| {\widetilde S}_{5  3  2} \big>+(\tau^{4} q^{2}+2 \tau^{3} q^{3}+\tau^{4} q^{3}+3 \tau^{2} q^{4}+3 \tau^{3} q^{4}+\tau q^{5}+4 \tau^{2} q^{5}+\tau^{3} q^{5}+2 \tau q^{6}+2 \tau^{2} q^{6}+\tau q^{7}) \big| {\widetilde S}_{5  3  1  1} \big>+(\tau^{3} q^{3}+\tau^{4} q^{3}+\tau^{2} q^{4}+3 \tau^{3} q^{4}+\tau^{4} q^{4}+3 \tau^{2} q^{5}+3 \tau^{3} q^{5}+\tau q^{6}+3 \tau^{2} q^{6}+\tau^{3} q^{6}+\tau q^{7}+\tau^{2} q^{7}) \big| {\widetilde S}_{5  2  2  1} \big>+(\tau^{3} q^{4}+\tau^{2} q^{5}+\tau^{3} q^{5}+\tau^{2} q^{6}) \big| {\widetilde S}_{5  2  1  1  1} \big>+(\tau^{4} q^{2}+\tau^{3} q^{3}+\tau^{4} q^{3}+2 \tau^{2} q^{4}+2 \tau^{3} q^{4}+\tau^{4} q^{4}+2 \tau^{2} q^{5}+\tau^{3} q^{5}+\tau q^{6}+2 \tau^{2} q^{6}+\tau q^{7}+q^{8}) \big| {\widetilde S}_{4  4  2} \big>+(\tau^{3} q^{3}+\tau^{4} q^{3}+2 \tau^{3} q^{4}+2 \tau^{2} q^{5}+\tau^{3} q^{5}+\tau q^{6}+\tau^{2} q^{6}+\tau q^{7}) \big| {\widetilde S}_{4  4  1  1} \big>+(\tau^{3} q^{3}+\tau^{4} q^{3}+2 \tau^{3} q^{4}+2 \tau^{2} q^{5}+\tau^{3} q^{5}+\tau q^{6}+\tau^{2} q^{6}+\tau q^{7}) \big| {\widetilde S}_{4  3  3} \big>+(\tau^{4} q^{3}+3 \tau^{3} q^{4}+2 \tau^{4} q^{4}+2 \tau^{2} q^{5}+5 \tau^{3} q^{5}+\tau^{4} q^{5}+4 \tau^{2} q^{6}+2 \tau^{3} q^{6}+\tau q^{7}+2 \tau^{2} q^{7}+\tau q^{8}) \big| {\widetilde S}_{4  3  2  1} \big>+(\tau^{4} q^{4}+\tau^{3} q^{5}+\tau^{2} q^{6}+\tau^{3} q^{6}+\tau^{2} q^{7}) \big| {\widetilde S}_{4  3  1  1  1} \big>+(\tau^{4} q^{4}+\tau^{3} q^{5}+\tau^{4} q^{5}+\tau^{2} q^{6}+2 \tau^{3} q^{6}+\tau^{4} q^{6}+\tau^{2} q^{7}+\tau^{3} q^{7}+\tau^{2} q^{8}) \big| {\widetilde S}_{4  2  2  2} \big>+(\tau^{3} q^{5}+\tau^{4} q^{5}+2 \tau^{3} q^{6}+\tau^{2} q^{7}+\tau^{3} q^{7}) \big| {\widetilde S}_{4  2  2  1  1} \big>+(\tau^{4} q^{4}+\tau^{3} q^{5}+\tau^{2} q^{6}+\tau^{3} q^{6}+\tau^{2} q^{7}) \big| {\widetilde S}_{3  3  3  1} \big>+(\tau^{3} q^{5}+\tau^{4} q^{5}+2 \tau^{3} q^{6}+\tau^{2} q^{7}+\tau^{3} q^{7}) \big| {\widetilde S}_{3  3  2  2} \big>+(\tau^{4} q^{5}+\tau^{3} q^{6}+\tau^{4} q^{6}+\tau^{3} q^{7}+\tau^{2} q^{8}) \big| {\widetilde S}_{3  3  2  1  1} \big>+(\tau^{4} q^{6}+\tau^{3} q^{7}+\tau^{4} q^{7}+\tau^{3} q^{8}) \big| {\widetilde S}_{3  2  2  2  1} \big>+\tau^{4} q^{8} \big| {\widetilde S}_{2  2  2  2  2} \big>
$
\end{center}

\begin{center}
$ \big| P^{(2,1)}_{[41]} \big> =
\big| {\widetilde S}_{10} \big>+(\tau+q +q^{2}+q^{3}) \big| {\widetilde S}_{9  1} \big>+(\tau^{2}+\tau q +q^{2}+\tau q^{2}+q^{3}+\tau q^{3}+2 q^{4}+q^{5}+q^{6}) \big| {\widetilde S}_{8  2} \big>+(\tau q +\tau q^{2}+q^{3}+\tau q^{3}+q^{4}+q^{5}) \big| {\widetilde S}_{8  1  1} \big>+(\tau^{2} q +\tau q^{2}+\tau^{2} q^{2}+q^{3}+2 \tau q^{3}+q^{4}+2 \tau q^{4}+2 q^{5}+\tau q^{5}+2 q^{6}+q^{7}) \big| {\widetilde S}_{7  3} \big>+(\tau^{2} q +\tau q^{2}+\tau^{2} q^{2}+2 \tau q^{3}+\tau^{2} q^{3}+q^{4}+3 \tau q^{4}+2 q^{5}+2 \tau q^{5}+2 q^{6}+\tau q^{6}+2 q^{7}+q^{8}) \big| {\widetilde S}_{7  2  1} \big>+(\tau q^{3}+\tau q^{4}+\tau q^{5}+q^{6}) \big| {\widetilde S}_{7  1  1  1} \big>+(\tau^{2} q^{2}+\tau q^{3}+\tau^{2} q^{3}+q^{4}+2 \tau q^{4}+\tau^{2} q^{4}+q^{5}+2 \tau q^{5}+2 q^{6}+\tau q^{6}+q^{7}+q^{8}) \big| {\widetilde S}_{6  4} \big>+(\tau^{2} q^{2}+\tau q^{3}+2 \tau^{2} q^{3}+3 \tau q^{4}+2 \tau^{2} q^{4}+q^{5}+5 \tau q^{5}+\tau^{2} q^{5}+2 q^{6}+4 \tau q^{6}+3 q^{7}+2 \tau q^{7}+2 q^{8}+q^{9}) \big| {\widetilde S}_{6  3  1} \big>+(\tau^{2} q^{2}+\tau^{2} q^{3}+\tau q^{4}+2 \tau^{2} q^{4}+2 \tau q^{5}+\tau^{2} q^{5}+q^{6}+2 \tau q^{6}+\tau^{2} q^{6}+q^{7}+2 \tau q^{7}+2 q^{8}+\tau q^{8}+q^{9}+q^{10}) \big| {\widetilde S}_{6  2  2} \big>+(\tau^{2} q^{3}+\tau q^{4}+\tau^{2} q^{4}+2 \tau q^{5}+\tau^{2} q^{5}+3 \tau q^{6}+q^{7}+2 \tau q^{7}+q^{8}+\tau q^{8}+q^{9}) \big| {\widetilde S}_{6  2  1  1} \big>+\tau q^{6} \big| {\widetilde S}_{6  1  1  1  1} \big>+(\tau^{2} q^{3}+\tau q^{4}+q^{5}+\tau q^{5}+\tau q^{6}+q^{7}) \big| {\widetilde S}_{5  5} \big>+(\tau^{2} q^{3}+\tau q^{4}+2 \tau^{2} q^{4}+3 \tau q^{5}+2 \tau^{2} q^{5}+q^{6}+4 \tau q^{6}+\tau^{2} q^{6}+2 q^{7}+3 \tau q^{7}+2 q^{8}+\tau q^{8}+q^{9}) \big| {\widetilde S}_{5  4  1} \big>+(\tau^{2} q^{3}+2 \tau^{2} q^{4}+2 \tau q^{5}+3 \tau^{2} q^{5}+4 \tau q^{6}+2 \tau^{2} q^{6}+q^{7}+5 \tau q^{7}+\tau^{2} q^{7}+2 q^{8}+3 \tau q^{8}+2 q^{9}+\tau q^{9}+q^{10}) \big| {\widetilde S}_{5  3  2} \big>+(\tau^{2} q^{4}+\tau q^{5}+2 \tau^{2} q^{5}+3 \tau q^{6}+2 \tau^{2} q^{6}+4 \tau q^{7}+\tau^{2} q^{7}+q^{8}+3 \tau q^{8}+q^{9}+\tau q^{9}+q^{10}) \big| {\widetilde S}_{5  3  1  1} \big>+(\tau^{2} q^{4}+2 \tau^{2} q^{5}+\tau q^{6}+2 \tau^{2} q^{6}+2 \tau q^{7}+2 \tau^{2} q^{7}+3 \tau q^{8}+\tau^{2} q^{8}+q^{9}+2 \tau q^{9}+q^{10}+\tau q^{10}+q^{11}) \big| {\widetilde S}_{5  2  2  1} \big>+(\tau^{2} q^{6}+\tau q^{7}+\tau q^{8}+\tau q^{9}) \big| {\widetilde S}_{5  2  1  1  1} \big>+(\tau^{2} q^{4}+\tau^{2} q^{5}+2 \tau q^{6}+2 \tau^{2} q^{6}+2 \tau q^{7}+\tau^{2} q^{7}+q^{8}+2 \tau q^{8}+\tau^{2} q^{8}+q^{9}+\tau q^{9}+q^{10}) \big| {\widetilde S}_{4  4  2} \big>+(\tau^{2} q^{5}+\tau q^{6}+\tau^{2} q^{6}+2 \tau q^{7}+\tau^{2} q^{7}+2 \tau q^{8}+q^{9}+\tau q^{9}) \big| {\widetilde S}_{4  4  1  1} \big>+(\tau^{2} q^{5}+\tau q^{6}+\tau^{2} q^{6}+2 \tau q^{7}+\tau^{2} q^{7}+2 \tau q^{8}+q^{9}+\tau q^{9}) \big| {\widetilde S}_{4  3  3} \big>+(\tau^{2} q^{5}+3 \tau^{2} q^{6}+2 \tau q^{7}+3 \tau^{2} q^{7}+4 \tau q^{8}+2 \tau^{2} q^{8}+4 \tau q^{9}+\tau^{2} q^{9}+q^{10}+2 \tau q^{10}+q^{11}) \big| {\widetilde S}_{4  3  2  1} \big>+(\tau^{2} q^{7}+\tau q^{8}+\tau^{2} q^{8}+\tau q^{9}+\tau q^{10}) \big| {\widetilde S}_{4  3  1  1  1} \big>+(\tau^{2} q^{6}+\tau^{2} q^{7}+2 \tau^{2} q^{8}+\tau q^{9}+\tau^{2} q^{9}+\tau q^{10}+\tau^{2} q^{10}+\tau q^{11}+q^{12}) \big| {\widetilde S}_{4  2  2  2} \big>+(\tau^{2} q^{7}+\tau^{2} q^{8}+\tau q^{9}+\tau^{2} q^{9}+\tau q^{10}+\tau q^{11}) \big| {\widetilde S}_{4  2  2  1  1} \big>+(\tau^{2} q^{7}+\tau q^{8}+\tau^{2} q^{8}+\tau q^{9}+\tau q^{10}) \big| {\widetilde S}_{3  3  3  1} \big>+(\tau^{2} q^{7}+\tau^{2} q^{8}+\tau q^{9}+\tau^{2} q^{9}+\tau q^{10}+\tau q^{11}) \big| {\widetilde S}_{3  3  2  2} \big>+(\tau^{2} q^{8}+\tau^{2} q^{9}+\tau q^{10}+\tau^{2} q^{10}+\tau q^{11}) \big| {\widetilde S}_{3  3  2  1  1} \big>+(\tau^{2} q^{9}+\tau^{2} q^{10}+\tau^{2} q^{11}+\tau q^{12}) \big| {\widetilde S}_{3  2  2  2  1} \big>+\tau^{2} q^{12} \big| {\widetilde S}_{2  2  2  2  2} \big>
$
\end{center}

\begin{center}
$ \big| P^{(2,1)}_{[5]} \big> =
\big| {\widetilde S}_{10} \big>+(q +q^{2}+q^{3}+q^{4}) \big| {\widetilde S}_{9  1} \big>+(q^{2}+q^{3}+2 q^{4}+2 q^{5}+2 q^{6}+q^{7}+q^{8}) \big| {\widetilde S}_{8  2} \big>+(q^{3}+q^{4}+2 q^{5}+q^{6}+q^{7}) \big| {\widetilde S}_{8  1  1} \big>+(q^{3}+q^{4}+2 q^{5}+3 q^{6}+3 q^{7}+2 q^{8}+2 q^{9}+q^{10}) \big| {\widetilde S}_{7  3} \big>+(q^{4}+2 q^{5}+3 q^{6}+4 q^{7}+4 q^{8}+3 q^{9}+2 q^{10}+q^{11}) \big| {\widetilde S}_{7  2  1} \big>+(q^{6}+q^{7}+q^{8}+q^{9}) \big| {\widetilde S}_{7  1  1  1} \big>+(q^{4}+q^{5}+2 q^{6}+2 q^{7}+3 q^{8}+2 q^{9}+2 q^{10}+q^{11}+q^{12}) \big| {\widetilde S}_{6  4} \big>+(q^{5}+2 q^{6}+4 q^{7}+5 q^{8}+6 q^{9}+5 q^{10}+4 q^{11}+2 q^{12}+q^{13}) \big| {\widetilde S}_{6  3  1} \big>+(q^{6}+q^{7}+3 q^{8}+3 q^{9}+4 q^{10}+3 q^{11}+3 q^{12}+q^{13}+q^{14}) \big| {\widetilde S}_{6  2  2} \big>+(q^{7}+2 q^{8}+3 q^{9}+3 q^{10}+3 q^{11}+2 q^{12}+q^{13}) \big| {\widetilde S}_{6  2  1  1} \big>+q^{10} \big| {\widetilde S}_{6  1  1  1  1} \big>+(q^{5}+q^{7}+q^{8}+q^{9}+q^{10}+q^{11}) \big| {\widetilde S}_{5  5} \big>+(q^{6}+2 q^{7}+3 q^{8}+4 q^{9}+4 q^{10}+4 q^{11}+3 q^{12}+2 q^{13}+q^{14}) \big| {\widetilde S}_{5  4  1} \big>+(q^{7}+2 q^{8}+4 q^{9}+5 q^{10}+6 q^{11}+5 q^{12}+4 q^{13}+2 q^{14}+q^{15}) \big| {\widetilde S}_{5  3  2} \big>+(q^{8}+2 q^{9}+4 q^{10}+4 q^{11}+4 q^{12}+3 q^{13}+2 q^{14}+q^{15}) \big| {\widetilde S}_{5  3  1  1} \big>+(q^{9}+2 q^{10}+3 q^{11}+4 q^{12}+4 q^{13}+3 q^{14}+2 q^{15}+q^{16}) \big| {\widetilde S}_{5  2  2  1} \big>+(q^{11}+q^{12}+q^{13}+q^{14}) \big| {\widetilde S}_{5  2  1  1  1} \big>+(q^{8}+q^{9}+3 q^{10}+2 q^{11}+3 q^{12}+2 q^{13}+2 q^{14}+q^{15}+q^{16}) \big| {\widetilde S}_{4  4  2} \big>+(q^{9}+q^{10}+2 q^{11}+2 q^{12}+2 q^{13}+q^{14}+q^{15}) \big| {\widetilde S}_{4  4  1  1} \big>+(q^{9}+q^{10}+2 q^{11}+2 q^{12}+2 q^{13}+q^{14}+q^{15}) \big| {\widetilde S}_{4  3  3} \big>+(q^{10}+3 q^{11}+4 q^{12}+5 q^{13}+5 q^{14}+3 q^{15}+2 q^{16}+q^{17}) \big| {\widetilde S}_{4  3  2  1} \big>+(q^{12}+q^{13}+q^{14}+q^{15}+q^{16}) \big| {\widetilde S}_{4  3  1  1  1} \big>+(q^{12}+q^{13}+2 q^{14}+2 q^{15}+2 q^{16}+q^{17}+q^{18}) \big| {\widetilde S}_{4  2  2  2} \big>+(q^{13}+q^{14}+2 q^{15}+q^{16}+q^{17}) \big| {\widetilde S}_{4  2  2  1  1} \big>+(q^{12}+q^{13}+q^{14}+q^{15}+q^{16}) \big| {\widetilde S}_{3  3  3  1} \big>+(q^{13}+q^{14}+2 q^{15}+q^{16}+q^{17}) \big| {\widetilde S}_{3  3  2  2} \big>+(q^{14}+q^{15}+q^{16}+q^{17}+q^{18}) \big| {\widetilde S}_{3  3  2  1  1} \big>+(q^{16}+q^{17}+q^{18}+q^{19}) \big| {\widetilde S}_{3  2  2  2  1} \big>+q^{20} \big| {\widetilde S}_{2  2  2  2  2} \big>$
\end{center} }

\subsection*{Series (3,3k+1)}

\subsubsection*{Fundamental representation}

\begin{align*}
\big| P^{(3,1)}_{[1]} \big> = \big| {\widetilde S}_{3} \big>
\end{align*}

\subsubsection*{Representations of size $|R| = 2$}

\begin{align*}
\big| P^{(3,1)}_{[11]} \big> =
\big| {\widetilde S}_{6} \big> + \tau \big| {\widetilde S}_{51} \big> + \tau^{2} \big| {\widetilde S}_{42} \big> + \tau^{3} \big| {\widetilde S}_{33} \big>
\end{align*}

\begin{align*}
\big| P^{(3,1)}_{[2]} \big> =
\big| {\widetilde S}_{6} \big> + q \big| {\widetilde S}_{51} \big> + q^2 \big| {\widetilde S}_{42} \big> + q^3 \big| {\widetilde S}_{33} \big>
\end{align*}

\subsubsection*{Representations of size $|R| = 3$}

\begin{center}
$ \big| P^{(3,1)}_{[111]} \big> = \big| {\widetilde S}_{9} \big>+(\tau+\tau^{2}) \big| {\widetilde S}_{8, 1} \big>+(\tau^{2}+\tau^{3}+\tau^{4}) \big| {\widetilde S}_{7, 2} \big>+\tau^{3} \big| {\widetilde S}_{7, 1, 1} \big>+(\tau^{3}+\tau^{4}+\tau^{5}+\tau^{6}) \big| {\widetilde S}_{6, 3} \big>+(\tau^{4}+\tau^{5}) \big| {\widetilde S}_{6, 2, 1} \big>+(\tau^{4}+\tau^{5}) \big| {\widetilde S}_{5, 4} \big>+(\tau^{5}+\tau^{6}+\tau^{7}) \big| {\widetilde S}_{5, 3, 1} \big>+\tau^{6} \big| {\widetilde S}_{5, 2, 2} \big>+\tau^{6} \big| {\widetilde S}_{4, 4, 1} \big>+(\tau^{7}+\tau^{8}) \big| {\widetilde S}_{4, 3, 2} \big>+\tau^{9} \big| {\widetilde S}_{3, 3, 3} \big>$
\end{center}

\begin{center}
$ \big| P^{(3,1)}_{[21]} \big> = \big| {\widetilde S}_{9} \big>+(\tau+q) \big| {\widetilde S}_{8, 1} \big>+(\tau^{2}+\tau q+q^{2}) \big| {\widetilde S}_{7, 2} \big>+\tau q \big| {\widetilde S}_{7, 1, 1} \big>+(\tau^{3}+\tau^{2} q+\tau q^{2}+q^{3}) \big| {\widetilde S}_{6, 3} \big>+(\tau^{2} q+\tau q^{2}) \big| {\widetilde S}_{6, 2, 1} \big>+(\tau^{2} q+\tau q^{2}) \big| {\widetilde S}_{5, 4} \big>+(\tau^{3} q+\tau^{2} q^{2}+\tau q^{3}) \big| {\widetilde S}_{5, 3, 1} \big>+\tau^{2} q^{2} \big| {\widetilde S}_{5, 2, 2} \big>+\tau^{2} q^{2} \big| {\widetilde S}_{4, 4, 1} \big>+(\tau^{3} q^{2}+\tau^{2} q^{3}) \big| {\widetilde S}_{4, 3, 2} \big>+\tau^{3} q^{3} \big| {\widetilde S}_{3, 3, 3} \big>$
\end{center}

\begin{center}
$ \big| P^{(3,1)}_{[3]} \big> = \big| {\widetilde S}_{9} \big>+(q+q^{2}) \big| {\widetilde S}_{8, 1} \big>+(q^{2}+q^{3}+q^{4}) \big| {\widetilde S}_{7, 2} \big>+q^{3} \big| {\widetilde S}_{7, 1, 1} \big>+(q^{3}+q^{4}+q^{5}+q^{6}) \big| {\widetilde S}_{6, 3} \big>+(q^{4}+q^{5}) \big| {\widetilde S}_{6, 2, 1} \big>+(q^{4}+q^{5}) \big| {\widetilde S}_{5, 4} \big>+(q^{5}+q^{6}+q^{7}) \big| {\widetilde S}_{5, 3, 1} \big>+q^{6} \big| {\widetilde S}_{5, 2, 2} \big>+q^{6} \big| {\widetilde S}_{4, 4, 1} \big>+(q^{7}+q^{8}) \big| {\widetilde S}_{4, 3, 2} \big>+q^{9} \big| {\widetilde S}_{3, 3, 3} \big>$
\end{center}

\subsubsection*{Representations of size $|R| = 4$}

{\fontsize{10pt}{0pt}

\begin{center}
$ \big| P^{(3,1)}_{[1111]} \big> =
\big| {\widetilde S}_{12} \big>+(\tau+\tau^{2}+\tau^{3}) \big| {\widetilde S}_{11, 1} \big>+(\tau^{2}+\tau^{3}+2 \tau^{4}+\tau^{5}+\tau^{6}) \big| {\widetilde S}_{10, 2} \big>+(\tau^{3}+\tau^{4}+\tau^{5}) \big| {\widetilde S}_{10, 1, 1} \big>+(\tau^{3}+\tau^{4}+2 \tau^{5}+2 \tau^{6}+2 \tau^{7}+\tau^{8}+\tau^{9}) \big| {\widetilde S}_{9  3} \big>+(\tau^{4}+2 \tau^{5}+2 \tau^{6}+2 \tau^{7}+\tau^{8}) \big| {\widetilde S}_{9  2  1} \big>+\tau^{6} \big| {\widetilde S}_{9  1  1  1} \big>+(\tau^{4}+\tau^{5}+2 \tau^{6}+2 \tau^{7}+3 \tau^{8}+\tau^{9}+\tau^{10}) \big| {\widetilde S}_{8  4} \big>+(\tau^{5}+2 \tau^{6}+3 \tau^{7}+3 \tau^{8}+3 \tau^{9}+2 \tau^{10}+\tau^{11}) \big| {\widetilde S}_{8  3  1} \big>+(\tau^{6}+\tau^{7}+2 \tau^{8}+\tau^{9}+\tau^{10}) \big| {\widetilde S}_{8  2  2} \big>+(\tau^{7}+\tau^{8}+\tau^{9}) \big| {\widetilde S}_{8  2  1  1} \big>+(\tau^{5}+\tau^{6}+2 \tau^{7}+\tau^{8}+2 \tau^{9}+\tau^{10}+\tau^{11}) \big| {\widetilde S}_{7  5} \big>+(\tau^{6}+2 \tau^{7}+3 \tau^{8}+4 \tau^{9}+3 \tau^{10}+2 \tau^{11}+\tau^{12}) \big| {\widetilde S}_{7  4  1} \big>+(\tau^{7}+2 \tau^{8}+3 \tau^{9}+3 \tau^{10}+3 \tau^{11}+2 \tau^{12}+\tau^{13}) \big| {\widetilde S}_{7  3  2} \big>+(\tau^{8}+\tau^{9}+2 \tau^{10}+\tau^{11}+\tau^{12}) \big| {\widetilde S}_{7  3  1  1} \big>+(\tau^{9}+\tau^{10}+\tau^{11}) \big| {\widetilde S}_{7  2  2  1} \big>+(\tau^{6}+\tau^{8}+\tau^{10}+\tau^{12}) \big| {\widetilde S}_{6  6} \big>+(\tau^{7}+2 \tau^{8}+2 \tau^{9}+2 \tau^{10}+2 \tau^{11}+\tau^{12}+\tau^{13}) \big| {\widetilde S}_{6  5  1} \big>+(\tau^{8}+2 \tau^{9}+4 \tau^{10}+3 \tau^{11}+3 \tau^{12}+\tau^{13}+\tau^{14}) \big| {\widetilde S}_{6  4  2} \big>+(\tau^{9}+\tau^{10}+2 \tau^{11}+\tau^{12}+\tau^{13}) \big| {\widetilde S}_{6  4  1  1} \big>+(\tau^{9}+\tau^{10}+2 \tau^{11}+2 \tau^{12}+2 \tau^{13}+\tau^{14}+\tau^{15}) \big| {\widetilde S}_{6  3  3} \big>+(\tau^{10}+2 \tau^{11}+2 \tau^{12}+2 \tau^{13}+\tau^{14}) \big| {\widetilde S}_{6  3  2  1} \big>+\tau^{12} \big| {\widetilde S}_{6  2  2  2} \big>+(\tau^{9}+\tau^{10}+2 \tau^{11}+\tau^{12}+\tau^{13}) \big| {\widetilde S}_{5  5  2} \big>+(\tau^{10}+\tau^{12}+\tau^{14}) \big| {\widetilde S}_{5  5  1  1} \big>+(\tau^{10}+2 \tau^{11}+2 \tau^{12}+2 \tau^{13}+\tau^{14}) \big| {\widetilde S}_{5  4  3} \big>+(\tau^{11}+2 \tau^{12}+2 \tau^{13}+\tau^{14}+\tau^{15}) \big| {\widetilde S}_{5  4  2  1} \big>+(\tau^{12}+\tau^{13}+2 \tau^{14}+\tau^{15}+\tau^{16}) \big| {\widetilde S}_{5  3  3  1} \big>+(\tau^{13}+\tau^{14}+\tau^{15}) \big| {\widetilde S}_{5  3  2  2} \big>+\tau^{12} \big| {\widetilde S}_{4  4  4} \big>+(\tau^{13}+\tau^{14}+\tau^{15}) \big| {\widetilde S}_{4  4  3  1} \big>+(\tau^{14}+\tau^{16}) \big| {\widetilde S}_{4  4  2  2} \big>+(\tau^{15}+\tau^{16}+\tau^{17}) \big| {\widetilde S}_{4  3  3  2} \big>+\tau^{18} \big| {\widetilde S}_{3  3  3  3} \big>$
\end{center}

\begin{center}
$ \big| P^{(3,1)}_{[211]} \big> = \big| {\widetilde S}_{12} \big>+(\tau+\tau^{2}+q) \big| {\widetilde S}_{11, 1} \big>+(\tau^{2}+\tau^{3}+\tau^{4}+\tau q+\tau^{2} q+q^{2}) \big| {\widetilde S}_{10, 2} \big>+(\tau^{3}+\tau q+\tau^{2} q) \big| {\widetilde S}_{10, 1, 1} \big>+(\tau^{3}+\tau^{4}+\tau^{5}+\tau^{6}+\tau^{2} q+\tau^{3} q+\tau^{4} q+\tau q^{2}+\tau^{2} q^{2}+q^{3}) \big| {\widetilde S}_{9  3} \big>+(\tau^{4}+\tau^{5}+\tau^{2} q+2 \tau^{3} q+\tau^{4} q+\tau q^{2}+\tau^{2} q^{2}) \big| {\widetilde S}_{9  2  1} \big>+\tau^{3} q \big| {\widetilde S}_{9  1  1  1} \big>+(\tau^{4}+\tau^{5}+\tau^{6}+\tau^{3} q+2 \tau^{4} q+\tau^{5} q+2 \tau^{2} q^{2}+\tau^{3} q^{2}+\tau q^{3}) \big| {\widetilde S}_{8  4} \big>+(\tau^{5}+\tau^{6}+\tau^{7}+\tau^{3} q+2 \tau^{4} q+2 \tau^{5} q+\tau^{6} q+\tau^{2} q^{2}+2 \tau^{3} q^{2}+\tau^{4} q^{2}+\tau q^{3}+\tau^{2} q^{3}) \big| {\widetilde S}_{8  3  1} \big>+(\tau^{6}+\tau^{4} q+\tau^{5} q+\tau^{2} q^{2}+\tau^{3} q^{2}+\tau^{4} q^{2}) \big| {\widetilde S}_{8  2  2} \big>+(\tau^{4} q+\tau^{5} q+\tau^{3} q^{2}) \big| {\widetilde S}_{8  2  1  1} \big>+(\tau^{5}+\tau^{6}+2 \tau^{4} q+\tau^{5} q+2 \tau^{3} q^{2}+\tau^{4} q^{2}+\tau^{2} q^{3}) \big| {\widetilde S}_{7  5} \big>+(\tau^{6}+\tau^{7}+\tau^{4} q+3 \tau^{5} q+2 \tau^{6} q+2 \tau^{3} q^{2}+3 \tau^{4} q^{2}+\tau^{5} q^{2}+\tau^{2} q^{3}+\tau^{3} q^{3}) \big| {\widetilde S}_{7  4  1} \big>+(\tau^{7}+\tau^{8}+\tau^{5} q+2 \tau^{6} q+\tau^{7} q+\tau^{3} q^{2}+2 \tau^{4} q^{2}+2 \tau^{5} q^{2}+\tau^{6} q^{2}+\tau^{2} q^{3}+\tau^{3} q^{3}+\tau^{4} q^{3}) \big| {\widetilde S}_{7  3  2} \big>+(\tau^{5} q+\tau^{6} q+\tau^{7} q+\tau^{4} q^{2}+\tau^{5} q^{2}+\tau^{3} q^{3}) \big| {\widetilde S}_{7  3  1  1} \big>+(\tau^{6} q+\tau^{4} q^{2}+\tau^{5} q^{2}) \big| {\widetilde S}_{7  2  2  1} \big>+(\tau^{6}+\tau^{5} q+\tau^{4} q^{2}+\tau^{3} q^{3}) \big| {\widetilde S}_{6  6} \big>+(\tau^{7}+2 \tau^{5} q+2 \tau^{6} q+2 \tau^{4} q^{2}+2 \tau^{5} q^{2}+\tau^{3} q^{3}+\tau^{4} q^{3}) \big| {\widetilde S}_{6  5  1} \big>+(\tau^{8}+2 \tau^{6} q+2 \tau^{7} q+2 \tau^{4} q^{2}+3 \tau^{5} q^{2}+2 \tau^{6} q^{2}+\tau^{3} q^{3}+\tau^{4} q^{3}+\tau^{5} q^{3}) \big| {\widetilde S}_{6  4  2} \big>+(\tau^{6} q+\tau^{7} q+2 \tau^{5} q^{2}+\tau^{6} q^{2}+\tau^{4} q^{3}) \big| {\widetilde S}_{6  4  1  1} \big>+(\tau^{9}+\tau^{7} q+\tau^{8} q+\tau^{5} q^{2}+\tau^{6} q^{2}+\tau^{7} q^{2}+\tau^{3} q^{3}+\tau^{4} q^{3}+\tau^{5} q^{3}+\tau^{6} q^{3}) \big| {\widetilde S}_{6  3  3} \big>+(\tau^{7} q+\tau^{8} q+\tau^{5} q^{2}+2 \tau^{6} q^{2}+\tau^{7} q^{2}+\tau^{4} q^{3}+\tau^{5} q^{3}) \big| {\widetilde S}_{6  3  2  1} \big>+\tau^{6} q^{2} \big| {\widetilde S}_{6  2  2  2} \big>+(\tau^{6} q+\tau^{7} q+2 \tau^{5} q^{2}+\tau^{6} q^{2}+\tau^{4} q^{3}) \big| {\widetilde S}_{5  5  2} \big>+(\tau^{7} q+\tau^{6} q^{2}+\tau^{5} q^{3}) \big| {\widetilde S}_{5  5  1  1} \big>+(\tau^{7} q+\tau^{8} q+\tau^{5} q^{2}+2 \tau^{6} q^{2}+\tau^{7} q^{2}+\tau^{4} q^{3}+\tau^{5} q^{3}) \big| {\widetilde S}_{5  4  3} \big>+(\tau^{8} q+2 \tau^{6} q^{2}+2 \tau^{7} q^{2}+\tau^{5} q^{3}+\tau^{6} q^{3}) \big| {\widetilde S}_{5  4  2  1} \big>+(\tau^{9} q+\tau^{7} q^{2}+\tau^{8} q^{2}+\tau^{5} q^{3}+\tau^{6} q^{3}+\tau^{7} q^{3}) \big| {\widetilde S}_{5  3  3  1} \big>+(\tau^{7} q^{2}+\tau^{8} q^{2}+\tau^{6} q^{3}) \big| {\widetilde S}_{5  3  2  2} \big>+\tau^{6} q^{2} \big| {\widetilde S}_{4  4  4} \big>+(\tau^{7} q^{2}+\tau^{8} q^{2}+\tau^{6} q^{3}) \big| {\widetilde S}_{4  4  3  1} \big>+(\tau^{8} q^{2}+\tau^{7} q^{3}) \big| {\widetilde S}_{4  4  2  2} \big>+(\tau^{9} q^{2}+\tau^{7} q^{3}+\tau^{8} q^{3}) \big| {\widetilde S}_{4  3  3  2} \big>+\tau^{9} q^{3} \big| {\widetilde S}_{3  3  3  3} \big>$
\end{center}

\begin{center}
$ \big| P^{(3,1)}_{[31]} \big> = \big| {\widetilde S}_{12} \big>+(\tau+q+q^{2}) \big| {\widetilde S}_{11, 1} \big>+(\tau^{2}+\tau q+q^{2}+\tau q^{2}+q^{3}+q^{4}) \big| {\widetilde S}_{10, 2} \big>+(\tau q+\tau q^{2}+q^{3}) \big| {\widetilde S}_{10, 1, 1} \big>+(\tau^{3}+\tau^{2} q+\tau q^{2}+\tau^{2} q^{2}+q^{3}+\tau q^{3}+q^{4}+\tau q^{4}+q^{5}+q^{6}) \big| {\widetilde S}_{9  3} \big>+(\tau^{2} q+\tau q^{2}+\tau^{2} q^{2}+2 \tau q^{3}+q^{4}+\tau q^{4}+q^{5}) \big| {\widetilde S}_{9  2  1} \big>+\tau q^{3} \big| {\widetilde S}_{9  1  1  1} \big>+(\tau^{3} q+2 \tau^{2} q^{2}+\tau q^{3}+\tau^{2} q^{3}+q^{4}+2 \tau q^{4}+q^{5}+\tau q^{5}+q^{6}) \big| {\widetilde S}_{8  4} \big>+(\tau^{3} q+\tau^{2} q^{2}+\tau^{3} q^{2}+\tau q^{3}+2 \tau^{2} q^{3}+2 \tau q^{4}+\tau^{2} q^{4}+q^{5}+2 \tau q^{5}+q^{6}+\tau q^{6}+q^{7}) \big| {\widetilde S}_{8  3  1} \big>+(\tau^{2} q^{2}+\tau^{2} q^{3}+\tau q^{4}+\tau^{2} q^{4}+\tau q^{5}+q^{6}) \big| {\widetilde S}_{8  2  2} \big>+(\tau^{2} q^{3}+\tau q^{4}+\tau q^{5}) \big| {\widetilde S}_{8  2  1  1} \big>+(\tau^{3} q^{2}+2 \tau^{2} q^{3}+2 \tau q^{4}+\tau^{2} q^{4}+q^{5}+\tau q^{5}+q^{6}) \big| {\widetilde S}_{7  5} \big>+(\tau^{3} q^{2}+2 \tau^{2} q^{3}+\tau^{3} q^{3}+\tau q^{4}+3 \tau^{2} q^{4}+3 \tau q^{5}+\tau^{2} q^{5}+q^{6}+2 \tau q^{6}+q^{7}) \big| {\widetilde S}_{7  4  1} \big>+(\tau^{3} q^{2}+\tau^{2} q^{3}+\tau^{3} q^{3}+2 \tau^{2} q^{4}+\tau^{3} q^{4}+\tau q^{5}+2 \tau^{2} q^{5}+2 \tau q^{6}+\tau^{2} q^{6}+q^{7}+\tau q^{7}+q^{8}) \big| {\widetilde S}_{7  3  2} \big>+(\tau^{3} q^{3}+\tau^{2} q^{4}+\tau q^{5}+\tau^{2} q^{5}+\tau q^{6}+\tau q^{7}) \big| {\widetilde S}_{7  3  1  1} \big>+(\tau^{2} q^{4}+\tau^{2} q^{5}+\tau q^{6}) \big| {\widetilde S}_{7  2  2  1} \big>+(\tau^{3} q^{3}+\tau^{2} q^{4}+\tau q^{5}+q^{6}) \big| {\widetilde S}_{6  6} \big>+(\tau^{3} q^{3}+2 \tau^{2} q^{4}+\tau^{3} q^{4}+2 \tau q^{5}+2 \tau^{2} q^{5}+2 \tau q^{6}+q^{7}) \big| {\widetilde S}_{6  5  1} \big>+(\tau^{3} q^{3}+2 \tau^{2} q^{4}+\tau^{3} q^{4}+3 \tau^{2} q^{5}+\tau^{3} q^{5}+2 \tau q^{6}+2 \tau^{2} q^{6}+2 \tau q^{7}+q^{8}) \big| {\widetilde S}_{6  4  2} \big>+(\tau^{3} q^{4}+2 \tau^{2} q^{5}+\tau q^{6}+\tau^{2} q^{6}+\tau q^{7}) \big| {\widetilde S}_{6  4  1  1} \big>+(\tau^{3} q^{3}+\tau^{3} q^{4}+\tau^{2} q^{5}+\tau^{3} q^{5}+\tau^{2} q^{6}+\tau^{3} q^{6}+\tau q^{7}+\tau^{2} q^{7}+\tau q^{8}+q^{9}) \big| {\widetilde S}_{6  3  3} \big>+(\tau^{3} q^{4}+\tau^{2} q^{5}+\tau^{3} q^{5}+2 \tau^{2} q^{6}+\tau q^{7}+\tau^{2} q^{7}+\tau q^{8}) \big| {\widetilde S}_{6  3  2  1} \big>+\tau^{2} q^{6} \big| {\widetilde S}_{6  2  2  2} \big>+(\tau^{3} q^{4}+2 \tau^{2} q^{5}+\tau q^{6}+\tau^{2} q^{6}+\tau q^{7}) \big| {\widetilde S}_{5  5  2} \big>+(\tau^{3} q^{5}+\tau^{2} q^{6}+\tau q^{7}) \big| {\widetilde S}_{5  5  1  1} \big>+(\tau^{3} q^{4}+\tau^{2} q^{5}+\tau^{3} q^{5}+2 \tau^{2} q^{6}+\tau q^{7}+\tau^{2} q^{7}+\tau q^{8}) \big| {\widetilde S}_{5  4  3} \big>+(\tau^{3} q^{5}+2 \tau^{2} q^{6}+\tau^{3} q^{6}+2 \tau^{2} q^{7}+\tau q^{8}) \big| {\widetilde S}_{5  4  2  1} \big>+(\tau^{3} q^{5}+\tau^{3} q^{6}+\tau^{2} q^{7}+\tau^{3} q^{7}+\tau^{2} q^{8}+\tau q^{9}) \big| {\widetilde S}_{5  3  3  1} \big>+(\tau^{3} q^{6}+\tau^{2} q^{7}+\tau^{2} q^{8}) \big| {\widetilde S}_{5  3  2  2} \big>+\tau^{2} q^{6} \big| {\widetilde S}_{4  4  4} \big>+(\tau^{3} q^{6}+\tau^{2} q^{7}+\tau^{2} q^{8}) \big| {\widetilde S}_{4  4  3  1} \big>+(\tau^{3} q^{7}+\tau^{2} q^{8}) \big| {\widetilde S}_{4  4  2  2} \big>+(\tau^{3} q^{7}+\tau^{3} q^{8}+\tau^{2} q^{9}) \big| {\widetilde S}_{4  3  3  2} \big>+\tau^{3} q^{9} \big| {\widetilde S}_{3  3  3  3} \big>$
\end{center}

\begin{center}
$ \big| P^{(3,1)}_{[4]} \big> = \big| {\widetilde S}_{12} \big>+(q+q^{2}+q^{3}) \big| {\widetilde S}_{11, 1} \big>+(q^{2}+q^{3}+2 q^{4}+q^{5}+q^{6}) \big| {\widetilde S}_{10, 2} \big>+(q^{3}+q^{4}+q^{5}) \big| {\widetilde S}_{10, 1, 1} \big>+(q^{3}+q^{4}+2 q^{5}+2 q^{6}+2 q^{7}+q^{8}+q^{9}) \big| {\widetilde S}_{9  3} \big>+(q^{4}+2 q^{5}+2 q^{6}+2 q^{7}+q^{8}) \big| {\widetilde S}_{9  2  1} \big>+q^{6} \big| {\widetilde S}_{9  1  1  1} \big>+(q^{4}+q^{5}+2 q^{6}+2 q^{7}+3 q^{8}+q^{9}+q^{10}) \big| {\widetilde S}_{8  4} \big>+(q^{5}+2 q^{6}+3 q^{7}+3 q^{8}+3 q^{9}+2 q^{10}+q^{11}) \big| {\widetilde S}_{8  3  1} \big>+(q^{6}+q^{7}+2 q^{8}+q^{9}+q^{10}) \big| {\widetilde S}_{8  2  2} \big>+(q^{7}+q^{8}+q^{9}) \big| {\widetilde S}_{8  2  1  1} \big>+(q^{5}+q^{6}+2 q^{7}+q^{8}+2 q^{9}+q^{10}+q^{11}) \big| {\widetilde S}_{7  5} \big>+(q^{6}+2 q^{7}+3 q^{8}+4 q^{9}+3 q^{10}+2 q^{11}+q^{12}) \big| {\widetilde S}_{7  4  1} \big>+(q^{7}+2 q^{8}+3 q^{9}+3 q^{10}+3 q^{11}+2 q^{12}+q^{13}) \big| {\widetilde S}_{7  3  2} \big>+(q^{8}+q^{9}+2 q^{10}+q^{11}+q^{12}) \big| {\widetilde S}_{7  3  1  1} \big>+(q^{9}+q^{10}+q^{11}) \big| {\widetilde S}_{7  2  2  1} \big>+(q^{6}+q^{8}+q^{10}+q^{12}) \big| {\widetilde S}_{6  6} \big>+(q^{7}+2 q^{8}+2 q^{9}+2 q^{10}+2 q^{11}+q^{12}+q^{13}) \big| {\widetilde S}_{6  5  1} \big>+(q^{8}+2 q^{9}+4 q^{10}+3 q^{11}+3 q^{12}+q^{13}+q^{14}) \big| {\widetilde S}_{6  4  2} \big>+(q^{9}+q^{10}+2 q^{11}+q^{12}+q^{13}) \big| {\widetilde S}_{6  4  1  1} \big>+(q^{9}+q^{10}+2 q^{11}+2 q^{12}+2 q^{13}+q^{14}+q^{15}) \big| {\widetilde S}_{6  3  3} \big>+(q^{10}+2 q^{11}+2 q^{12}+2 q^{13}+q^{14}) \big| {\widetilde S}_{6  3  2  1} \big>+q^{12} \big| {\widetilde S}_{6  2  2  2} \big>+(q^{9}+q^{10}+2 q^{11}+q^{12}+q^{13}) \big| {\widetilde S}_{5  5  2} \big>+(q^{10}+q^{12}+q^{14}) \big| {\widetilde S}_{5  5  1  1} \big>+(q^{10}+2 q^{11}+2 q^{12}+2 q^{13}+q^{14}) \big| {\widetilde S}_{5  4  3} \big>+(q^{11}+2 q^{12}+2 q^{13}+q^{14}+q^{15}) \big| {\widetilde S}_{5  4  2  1} \big>+(q^{12}+q^{13}+2 q^{14}+q^{15}+q^{16}) \big| {\widetilde S}_{5  3  3  1} \big>+(q^{13}+q^{14}+q^{15}) \big| {\widetilde S}_{5  3  2  2} \big>+q^{12} \big| {\widetilde S}_{4  4  4} \big>+(q^{13}+q^{14}+q^{15}) \big| {\widetilde S}_{4  4  3  1} \big>+(q^{14}+q^{16}) \big| {\widetilde S}_{4  4  2  2} \big>+(q^{15}+q^{16}+q^{17}) \big| {\widetilde S}_{4  3  3  2} \big>+q^{18} \big| {\widetilde S}_{3  3  3  3} \big>$
\end{center}

\begin{center}
$ \big| P^{(3,1)}_{[22]} \big> = \big| {\widetilde S}_{12} \big>+(\tau+q+\tau q) \big| {\widetilde S}_{11, 1} \big>+(\tau^{2}+\tau q+\tau^{2} q+q^{2}+\tau q^{2}+\tau^{2} q^{2}) \big| {\widetilde S}_{10, 2} \big>+(\tau q+\tau^{2} q+\tau q^{2}) \big| {\widetilde S}_{10, 1, 1} \big>+(\tau^{3}+\tau^{2} q+\tau^{3} q+\tau q^{2}+\tau^{2} q^{2}+\tau^{3} q^{2}+q^{3}+\tau q^{3}+\tau^{2} q^{3}+\tau^{3} q^{3}) \big| {\widetilde S}_{9  3} \big>+(\tau^{2} q+\tau^{3} q+\tau q^{2}+2 \tau^{2} q^{2}+\tau^{3} q^{2}+\tau q^{3}+\tau^{2} q^{3}) \big| {\widetilde S}_{9  2  1} \big>+\tau^{2} q^{2} \big| {\widetilde S}_{9  1  1  1} \big>+(\tau^{4}+\tau^{3} q+\tau^{4} q+\tau^{2} q^{2}+\tau^{3} q^{2}+\tau^{4} q^{2}+\tau q^{3}+\tau^{2} q^{3}+q^{4}+\tau q^{4}+\tau^{2} q^{4}) \big| {\widetilde S}_{8  4} \big>+(\tau^{3} q+\tau^{4} q+\tau^{2} q^{2}+2 \tau^{3} q^{2}+\tau^{4} q^{2}+\tau q^{3}+2 \tau^{2} q^{3}+2 \tau^{3} q^{3}+\tau^{4} q^{3}+\tau q^{4}+\tau^{2} q^{4}+\tau^{3} q^{4}) \big| {\widetilde S}_{8  3  1} \big>+(\tau^{2} q^{2}+\tau^{3} q^{2}+\tau^{4} q^{2}+\tau^{2} q^{3}+\tau^{3} q^{3}+\tau^{2} q^{4}) \big| {\widetilde S}_{8  2  2} \big>+(\tau^{3} q^{2}+\tau^{2} q^{3}+\tau^{3} q^{3}) \big| {\widetilde S}_{8  2  1  1} \big>+(\tau^{5}+\tau^{4} q+\tau^{5} q+\tau^{3} q^{2}+\tau^{2} q^{3}+\tau^{3} q^{3}+\tau q^{4}+q^{5}+\tau q^{5}) \big| {\widetilde S}_{7  5} \big>+(\tau^{4} q+\tau^{5} q+\tau^{3} q^{2}+2 \tau^{4} q^{2}+\tau^{5} q^{2}+\tau^{2} q^{3}+2 \tau^{3} q^{3}+\tau^{4} q^{3}+\tau q^{4}+2 \tau^{2} q^{4}+\tau^{3} q^{4}+\tau q^{5}+\tau^{2} q^{5}) \big| {\widetilde S}_{7  4  1} \big>+(\tau^{3} q^{2}+\tau^{4} q^{2}+\tau^{5} q^{2}+\tau^{2} q^{3}+2 \tau^{3} q^{3}+2 \tau^{4} q^{3}+\tau^{5} q^{3}+\tau^{2} q^{4}+2 \tau^{3} q^{4}+\tau^{4} q^{4}+\tau^{2} q^{5}+\tau^{3} q^{5}) \big| {\widetilde S}_{7  3  2} \big>+(\tau^{4} q^{2}+\tau^{3} q^{3}+\tau^{4} q^{3}+\tau^{2} q^{4}+\tau^{3} q^{4}+\tau^{4} q^{4}) \big| {\widetilde S}_{7  3  1  1} \big>+(\tau^{3} q^{3}+\tau^{4} q^{3}+\tau^{3} q^{4}) \big| {\widetilde S}_{7  2  2  1} \big>+(\tau^{6}+\tau^{4} q^{2}+\tau^{2} q^{4}+q^{6}) \big| {\widetilde S}_{6  6} \big>+(\tau^{5} q+\tau^{6} q+\tau^{4} q^{2}+\tau^{5} q^{2}+\tau^{3} q^{3}+\tau^{4} q^{3}+\tau^{2} q^{4}+\tau^{3} q^{4}+\tau q^{5}+\tau^{2} q^{5}+\tau q^{6}) \big| {\widetilde S}_{6  5  1} \big>+(\tau^{4} q^{2}+\tau^{5} q^{2}+\tau^{6} q^{2}+\tau^{3} q^{3}+2 \tau^{4} q^{3}+\tau^{5} q^{3}+\tau^{2} q^{4}+2 \tau^{3} q^{4}+2 \tau^{4} q^{4}+\tau^{2} q^{5}+\tau^{3} q^{5}+\tau^{2} q^{6}) \big| {\widetilde S}_{6  4  2} \big>+(\tau^{5} q^{2}+\tau^{4} q^{3}+\tau^{5} q^{3}+\tau^{3} q^{4}+\tau^{2} q^{5}+\tau^{3} q^{5}) \big| {\widetilde S}_{6  4  1  1} \big>+(\tau^{3} q^{3}+\tau^{4} q^{3}+\tau^{5} q^{3}+\tau^{6} q^{3}+\tau^{3} q^{4}+\tau^{4} q^{4}+\tau^{5} q^{4}+\tau^{3} q^{5}+\tau^{4} q^{5}+\tau^{3} q^{6}) \big| {\widetilde S}_{6  3  3} \big>+(\tau^{4} q^{3}+\tau^{5} q^{3}+\tau^{3} q^{4}+2 \tau^{4} q^{4}+\tau^{5} q^{4}+\tau^{3} q^{5}+\tau^{4} q^{5}) \big| {\widetilde S}_{6  3  2  1} \big>+\tau^{4} q^{4} \big| {\widetilde S}_{6  2  2  2} \big>+(\tau^{5} q^{2}+\tau^{4} q^{3}+\tau^{5} q^{3}+\tau^{3} q^{4}+\tau^{2} q^{5}+\tau^{3} q^{5}) \big| {\widetilde S}_{5  5  2} \big>+(\tau^{6} q^{2}+\tau^{4} q^{4}+\tau^{2} q^{6}) \big| {\widetilde S}_{5  5  1  1} \big>+(\tau^{4} q^{3}+\tau^{5} q^{3}+\tau^{3} q^{4}+2 \tau^{4} q^{4}+\tau^{5} q^{4}+\tau^{3} q^{5}+\tau^{4} q^{5}) \big| {\widetilde S}_{5  4  3} \big>+(\tau^{5} q^{3}+\tau^{6} q^{3}+\tau^{4} q^{4}+\tau^{5} q^{4}+\tau^{3} q^{5}+\tau^{4} q^{5}+\tau^{3} q^{6}) \big| {\widetilde S}_{5  4  2  1} \big>+(\tau^{4} q^{4}+\tau^{5} q^{4}+\tau^{6} q^{4}+\tau^{4} q^{5}+\tau^{5} q^{5}+\tau^{4} q^{6}) \big| {\widetilde S}_{5  3  3  1} \big>+(\tau^{5} q^{4}+\tau^{4} q^{5}+\tau^{5} q^{5}) \big| {\widetilde S}_{5  3  2  2} \big>+\tau^{4} q^{4} \big| {\widetilde S}_{4  4  4} \big>+(\tau^{5} q^{4}+\tau^{4} q^{5}+\tau^{5} q^{5}) \big| {\widetilde S}_{4  4  3  1} \big>+(\tau^{6} q^{4}+\tau^{4} q^{6}) \big| {\widetilde S}_{4  4  2  2} \big>+(\tau^{5} q^{5}+\tau^{6} q^{5}+\tau^{5} q^{6}) \big| {\widetilde S}_{4  3  3  2} \big>+\tau^{6} q^{6} \big| {\widetilde S}_{3  3  3  3} \big>$
\end{center} }

\subsection*{Series (4,4k+1)}

\subsubsection*{Fundamental representation}

\begin{align*}
\big| P^{(4,1)}_{[1]} \big> = \big| {\widetilde S}_{4} \big>
\end{align*}

\subsubsection*{Representations of size $|R| = 2$}

\begin{align*}
\big| P^{(4,1)}_{[11]} \big> =
\big| {\widetilde S}_{8} \big> + \tau \big| {\widetilde S}_{71} \big> + \tau^{2} \big| {\widetilde S}_{62} \big> + \tau^{3} \big| {\widetilde S}_{53} \big> + \tau^{4} \big| {\widetilde S}_{44} \big> \\
\big| P^{(4,1)}_{[2]} \big> =
\big| {\widetilde S}_{8} \big> + q \big| {\widetilde S}_{71} \big> + q^{2} \big| {\widetilde S}_{62} \big> + q^{3} \big| {\widetilde S}_{53} \big> + q^{4} \big| {\widetilde S}_{44} \big>
\end{align*}

\subsubsection*{Representations of size $|R| = 3$}

{\fontsize{8pt}{0pt}

\begin{center}
$\big| P^{(4, 1)}_{[111]} \big> = \big| {\widetilde S}_{12} \big>+(\tau+\tau^{2}) \big| {\widetilde S}_{11, 1} \big>+(\tau^{2}+\tau^{3}+\tau^{4}) \big| {\widetilde S}_{10, 2} \big>+\tau^{3} \big| {\widetilde S}_{10, 1, 1} \big>+(\tau^{3}+\tau^{4}+\tau^{5}+\tau^{6}) \big| {\widetilde S}_{9  3} \big>+(\tau^{4}+\tau^{5}) \big| {\widetilde S}_{9  2  1} \big>+(\tau^{4}+\tau^{5}+\tau^{6}+\tau^{7}+\tau^{8}) \big| {\widetilde S}_{8  4} \big>+(\tau^{5}+\tau^{6}+\tau^{7}) \big| {\widetilde S}_{8  3  1} \big>+\tau^{6} \big| {\widetilde S}_{8  2  2} \big>+(\tau^{5}+\tau^{6}+\tau^{7}) \big| {\widetilde S}_{7  5} \big>+(\tau^{6}+\tau^{7}+\tau^{8}+\tau^{9}) \big| {\widetilde S}_{7  4  1} \big>+(\tau^{7}+\tau^{8}) \big| {\widetilde S}_{7  3  2} \big>+\tau^{6} \big| {\widetilde S}_{6  6} \big>+(\tau^{7}+\tau^{8}) \big| {\widetilde S}_{6  5  1} \big>+(\tau^{8}+\tau^{9}+\tau^{10}) \big| {\widetilde S}_{6  4  2} \big>+\tau^{9} \big| {\widetilde S}_{6  3  3} \big>+\tau^{9} \big| {\widetilde S}_{5  5  2} \big>+(\tau^{10}+\tau^{11}) \big| {\widetilde S}_{5  4  3} \big>+\tau^{12} \big| {\widetilde S}_{4  4  4} \big>$
\end{center}

\begin{center}
$\big| P^{(4, 1)}_{[21]} \big> = \big| {\widetilde S}_{12} \big>+(\tau+q) \big| {\widetilde S}_{11, 1} \big>+(\tau^{2}+\tau q+q^{2}) \big| {\widetilde S}_{10, 2} \big>+\tau q \big| {\widetilde S}_{10, 1, 1} \big>+(\tau^{3}+\tau^{2} q+\tau q^{2}+q^{3}) \big| {\widetilde S}_{9  3} \big>+(\tau^{2} q+\tau q^{2}) \big| {\widetilde S}_{9  2  1} \big>+(\tau^{4}+\tau^{3} q+\tau^{2} q^{2}+\tau q^{3}+q^{4}) \big| {\widetilde S}_{8  4} \big>+(\tau^{3} q+\tau^{2} q^{2}+\tau q^{3}) \big| {\widetilde S}_{8  3  1} \big>+\tau^{2} q^{2} \big| {\widetilde S}_{8  2  2} \big>+(\tau^{3} q+\tau^{2} q^{2}+\tau q^{3}) \big| {\widetilde S}_{7  5} \big>+(\tau^{4} q+\tau^{3} q^{2}+\tau^{2} q^{3}+\tau q^{4}) \big| {\widetilde S}_{7  4  1} \big>+(\tau^{3} q^{2}+\tau^{2} q^{3}) \big| {\widetilde S}_{7  3  2} \big>+\tau^{2} q^{2} \big| {\widetilde S}_{6  6} \big>+(\tau^{3} q^{2}+\tau^{2} q^{3}) \big| {\widetilde S}_{6  5  1} \big>+(\tau^{4} q^{2}+\tau^{3} q^{3}+\tau^{2} q^{4}) \big| {\widetilde S}_{6  4  2} \big>+\tau^{3} q^{3} \big| {\widetilde S}_{6  3  3} \big>+\tau^{3} q^{3} \big| {\widetilde S}_{5  5  2} \big>+(\tau^{4} q^{3}+\tau^{3} q^{4}) \big| {\widetilde S}_{5  4  3} \big>+\tau^{4} q^{4} \big| {\widetilde S}_{4  4  4} \big>$
\end{center}

\begin{center}
$\big| P^{(4, 1)}_{[3]} \big> = \big| {\widetilde S}_{12} \big>+(q+q^{2}) \big| {\widetilde S}_{11, 1} \big>+(q^{2}+q^{3}+q^{4}) \big| {\widetilde S}_{10, 2} \big>+q^{3} \big| {\widetilde S}_{10, 1, 1} \big>+(q^{3}+q^{4}+q^{5}+q^{6}) \big| {\widetilde S}_{9  3} \big>+(q^{4}+q^{5}) \big| {\widetilde S}_{9  2  1} \big>+(q^{4}+q^{5}+q^{6}+q^{7}+q^{8}) \big| {\widetilde S}_{8  4} \big>+(q^{5}+q^{6}+q^{7}) \big| {\widetilde S}_{8  3  1} \big>+q^{6} \big| {\widetilde S}_{8  2  2} \big>+(q^{5}+q^{6}+q^{7}) \big| {\widetilde S}_{7  5} \big>+(q^{6}+q^{7}+q^{8}+q^{9}) \big| {\widetilde S}_{7  4  1} \big>+(q^{7}+q^{8}) \big| {\widetilde S}_{7  3  2} \big>+q^{6} \big| {\widetilde S}_{6  6} \big>+(q^{7}+q^{8}) \big| {\widetilde S}_{6  5  1} \big>+(q^{8}+q^{9}+q^{10}) \big| {\widetilde S}_{6  4  2} \big>+q^{9} \big| {\widetilde S}_{6  3  3} \big>+q^{9} \big| {\widetilde S}_{5  5  2} \big>+(q^{10}+q^{11}) \big| {\widetilde S}_{5  4  3} \big>+q^{12} \big| {\widetilde S}_{4  4  4} \big>$
\end{center} }


\begin{thebibliography}{100}

\bibitem{ChernSimons} S.-S.Chern and J. Simons, \emph{Characteristic forms and geometric invariants}, Annals of Mathematics 99-1 (1974) 48Ц69

\bibitem{WittenJones}  E.Witten, \emph{Quantum Field Theory and the Jones Polynomial}, Comm.Math.Phys.{\bf 121} (1989) 351

\bibitem{WittenLectures} S. Hu and E.Witten, \emph{Lecture Notes on Chern-Simons-Witten Theory}, World Scientific ( 2001)

\bibitem{Jones} V.Jones, \emph{Index for subfactors}, Invent. Math. 72 (1983) 1-25; \emph{A polynomial invariant for knots via von Neumann algebras}, Bull. Amer. Math. Soc. 12 (1985) 103-112; \emph{Hecke algebra representations of braid groups and link polynomials}, Ann. of Math. 126-2 (1987) 335-388

\bibitem{Homfly} Freyd, P.; Yetter, D., Hoste, J., Lickorish, W.B.R., Millett, K., and Ocneanu, A. \emph{A New Polynomial Invariant of Knots and Links}, Bulletin of the American Mathematical Society 12-2 (1985) 239Ц246

\bibitem{LecturesRCFT} G. Moore and N. Seiberg, \emph{Lectures on RCFT}, Physics, geometry, and topology (Banff, AB, 1989), ed. H.C. Lee, NATO Adv. Sci. Inst. Ser. B Phys., 238, Plenum, NY (1990) 263-361

\bibitem{LecturesCFT} P. Di Francesco, P. Mathieu and D. Senechal, \emph{Conformal field theory}, Springer 1997

\bibitem{Verlinde} E.Verlinde, \emph{Fusion Rules and Modular Transformations in 2D Conformal Field Theory}, Nucl. Phys. {\bf B300} (1988) 360

\bibitem{AKMV}
M.Aganagic, A.Klemm, M.Marino and C.Vafa,
"Matrix Model as a Mirror of Chern-Simons Theory",
JHEP 0402 (2004) 010, hep-th/0211098

\bibitem{Marino} M.Marino, \emph{Chern-Simons Theory, Matrix Models, and Topological Strings}, International Series of Monographs on Physics, Oxford Science Publications (2005)

\bibitem{RT} N. Reshetikhin, and V. Turaev, \emph{Invariants of 3-manifolds via link polynomials and quantum groups},  Invent. Math. 103-1 (1991) 547

\bibitem{LecturesQG} D.Freed, \emph{Quantum Groups from Path Integrals}, lecture notes for CRM-CAP Summer School ``Particles and Fields (1994), arXiv:q-alg/9501025

\bibitem{Khovanov1}
M. Khovanov, \emph{A Categorification Of The Jones Polynomial}, Duke. Math. J. 101 (2000) 359-426,  arXiv:math/9908171

\bibitem{Khovanov2}
M. Khovanov, \emph{Categorifications Of The Colored Jones Polynomial}, , J. Knot Theory Ramifications 14 (2005) 111Ц130, arXiv:math/0302060

\bibitem{KhovanovRozansky} Khovanov, M. and Rozansky, L., \emph{Matrix factorizations and link homology}, Fundamenta Mathematicae, vol.199 (2008), 1-91, arXiv:math/0401268; \emph{Matrix factorizations and link homology II}, Geometry and Topology, vol.12 (2008), 1387-1425, arXiv:math/0505056

\bibitem{Superpolynomial}
N. M. Dunfield, S. Gukov, and J. Rasmussen, \emph{The Superpolynomial For Knot Homologies}, Experiment. Math. 15 (2006) 129, math/0505662

\bibitem{AS} M.Aganagic and S.Shakirov,
  \emph{Knot Homology from Refined Chern-Simons Theory},
arXiv:1105.5117

\bibitem{MTC} A.Kirillov,Jr., \emph{On inner product in modular tensor categories. I}, arXiv:q-alg/9508017; \emph{On inner product in modular tensor categories. II. Inner product on conformal blocks and affine inner product identities}, arXiv:q-alg/9611008, Adv.Theor.Math.Phys.2:155-180,1998

\bibitem{AS2} M.Aganagic and S.Shakirov,
  \emph{Refined Chern-Simons Theory and Knot Homology}, Proc. String-Math Conference (2011), arXiv:1202.2489

\bibitem{MoscowHallLittlewood} A. Mironov, A. Morozov, Sh. Shakirov and A. Sleptsov, \emph{Interplay between MacDonald and Hall-Littlewood expansions of extended torus superpolynomials}, JHEP 2012 (2012) 70, arXiv:1201.3339

\bibitem{Cherednik} I.Cherednik, \emph{Jones polynomials of torus knots via DAHA}, arXiv:1111.6195

\bibitem{Quadr} E. Gorsky, S.Gukov and M. Stosic, \emph{Quadruply-graded colored homology of knots}, arXiv:1304.3481

\bibitem{Zexpand} S.Arthamonov, A.Mironov, A.Morozov, \emph{Differential hierarchy and additional grading of knot polynomials}, arXiv:1306.5682

\bibitem{GammaSmirnov} P. Dunin-Barkowski, A. Mironov, A. Morozov, A. Sleptsov and A. Smirnov, \emph{Superpolynomials for toric knots from evolution induced by cut-and-join operators}, JHEP 03 (2013) 021, arXiv:1106.4305

\bibitem{GammaOblomkov} A. Oblomkov, J. Rasmussen and V. Shende, \emph{The Hilbert scheme of a plane curve singularity and the HOMFLY homology of its link}, arXiv:1201.2115

\bibitem{GammaNegut} E.Gorsky and A.Negut, \emph{Refined knot invariants and Hilbert schemes}, arXiv:1304.3328

\bibitem{GammaGukov} H. Fuji, S.Gukov and P.Sulkowski, \emph{Volume Conjecture: Refined and Categorified}, arXiv:1203.2182

\bibitem{RossoJones} M.Rosso and V.Jones, \emph{On the invariants of torus knots derived from quantum groups}, J. Knot Theory Ramifications 2-1 (1993) 97Ц112

\bibitem{MoscowHallLittlewood2} A.Mironov, A.Morozov and Sh.Shakirov, \emph{Torus HOMFLY as the Hall-Littlewood Polynomials}, J. Phys. A: Math. Theor. 45 (2012) 355202, arXiv:1203.0667

\bibitem{HallLittlewoodGorsky} P. Etingof, E. Gorsky and I. Losev, \emph{Representations of Rational Cherednik algebras with minimal support and torus knots}, arXiv:1304.3412

\bibitem{BergeronGarsia1} F. Bergeron and G. Garsia, \emph{Science Fiction and Macdonald's Polynomials}, CRM Proceedings and Lecture Notes, American Mathematical Society (1998) 363-429, arXiv:math/9809128

\bibitem{BergeronGarsia2} F. Bergeron, A. Garsia, M. Haiman and G. Tesler, \emph{Identities and Positivity Conjectures for some remarkable Operators in the Theory of Symmetric Functions}, Theory of Symmetric Functions, Methods and Applications of Analysis (1999) 363-420

\bibitem{Macdonald} I.G. Macdonald, \emph{Symmetric Functions and Hall Polynomials}, 2nd ed. Oxford University Press, New York. 1999.

\bibitem{BorodinCorwin} A.Borodin and I.Corwin, \emph{Macdonald processes}, arXiv:1111.4408

\end{thebibliography}
\end{document}